\documentclass[journal]{IEEEtran}

\usepackage{amsmath}
\usepackage{amssymb}
\usepackage{bm}
\usepackage{color}
\usepackage{graphicx}

\usepackage[english]{babel}

\usepackage{graphicx}

\usepackage{tikz}
\usetikzlibrary{arrows,backgrounds}

\usepackage{pgfplots}
\pgfplotsset{compat=newest}

\newcommand{\sortbib}[1]{}

\usepackage{dsfont}

\newcommand{\di}{\ensuremath{\:\mathrm{d}}}                       
\newcommand{\T}{\ensuremath{\mathrm{T}}}                          

\newcommand{\dt}{\ensuremath{\tau}}
\newcommand{\dtref}{\ensuremath{\tau_{\text{\rm ref}}}}
\newcommand{\vref}{\ensuremath{v_{\text{\rm ref}}}}

\title{Cyber-physical Control of Road Freight Transport}
\author{B.~Besselink, V.~Turri, S.H.~van de Hoef, K.-Y.~Liang, A.~Alam, J.~M{\aa}rtensson, K.H.~Johansson%
  \thanks{B.~Besselink, V.~Turri, S.H.~van de Hoef, K.-Y.~Liang, J.~M{\aa}rtensson, and K.H.~Johansson are with the ACCESS Linnaeus Centre and the Department of Automatic Control, School of Electrical Engineering, KTH Royal Institute of Technology, Stockholm, Sweden. Email: bart.besselink@ee.kth.se, turri@kth.se, shvdh@kth.se, kyliang@kth.se, jonas.martensson@ee.kth.se, kallej@kth.se.}%
  \thanks{K.-Y.~Liang and A.~Alam are with Scania CV AB, S\"{o}dert\"{a}lje, Sweden. Email: assad.alam@scania.com.}}

\begin{document}

\maketitle

\begin{abstract}
Freight transportation is of outmost importance for our society and is continuously increasing. At the same time, transporting goods on roads accounts for about 26\% of all energy consumption and 18\% of all greenhouse gas emissions in the European Union. Despite the influence the transportation system has on our energy consumption and the environment, road transportation is mainly done by individual long-haulage trucks with no real-time coordination or global optimization. In this paper, we review how modern information and communication technology supports a cyber-physical transportation system architecture with an integrated logistic system coordinating fleets of trucks traveling together in vehicle platoons. From the reduced air drag, platooning trucks traveling close together can save about 10\% of their fuel consumption. Utilizing road grade information and vehicle-to-vehicle communication, a safe and fuel-optimized cooperative look-ahead control strategy is implemented on top of the existing cruise controller. By optimizing the interaction between vehicles and platoons of vehicles, it is shown that significant improvements can be achieved. An integrated transport planning and vehicle routing in the fleet management system allows both small and large fleet owners to benefit from the collaboration. A realistic case study with 200~heavy-duty vehicles performing transportation tasks in Sweden is described. Simulations show overall fuel savings at more than 5\% thanks to coordinated platoon planning. It is also illustrated how well the proposed cooperative look-ahead controller for heavy-duty vehicle platoons manages to optimize the velocity profiles of the vehicles over a hilly segment of the considered road network.
\end{abstract}

\section{Introduction}\label{sec_introduction}
The freight transportation sector is of great importance to our society and the demand for transportation is strongly linked to economic development. As a result of the growing world economy, the road freight transportation sector in the OECD in 2050 is projected to have grown by roughly $90$\% with respect to 2010 levels, according to a prediction of the International Transport Forum \cite{itf_transportoutlook_2015}. For developing countries, a significantly larger growth is expected. At the same time, the transportation sector is responsible for a large part of the world's energy consumption and (greenhouse gas) emissions. As an example, in 2012, the road transportation sector amounted for 26\% of the total energy consumption and 18\% of all greenhouse gas emissions in the European Union \cite{eu_pocketbook_2014}. This impact on the environment provides a strong motivation for developing a more fuel-efficient freight transportation sector, which is further encouraged by the fact that about one third of the cost of operating a heavy-duty vehicle is associated to its fuel consumption \cite{scania_annualreport_2014}.

\begin{figure}
\begin{center}
  \includegraphics[width=0.95\columnwidth,bb=0 0 1400 1000]{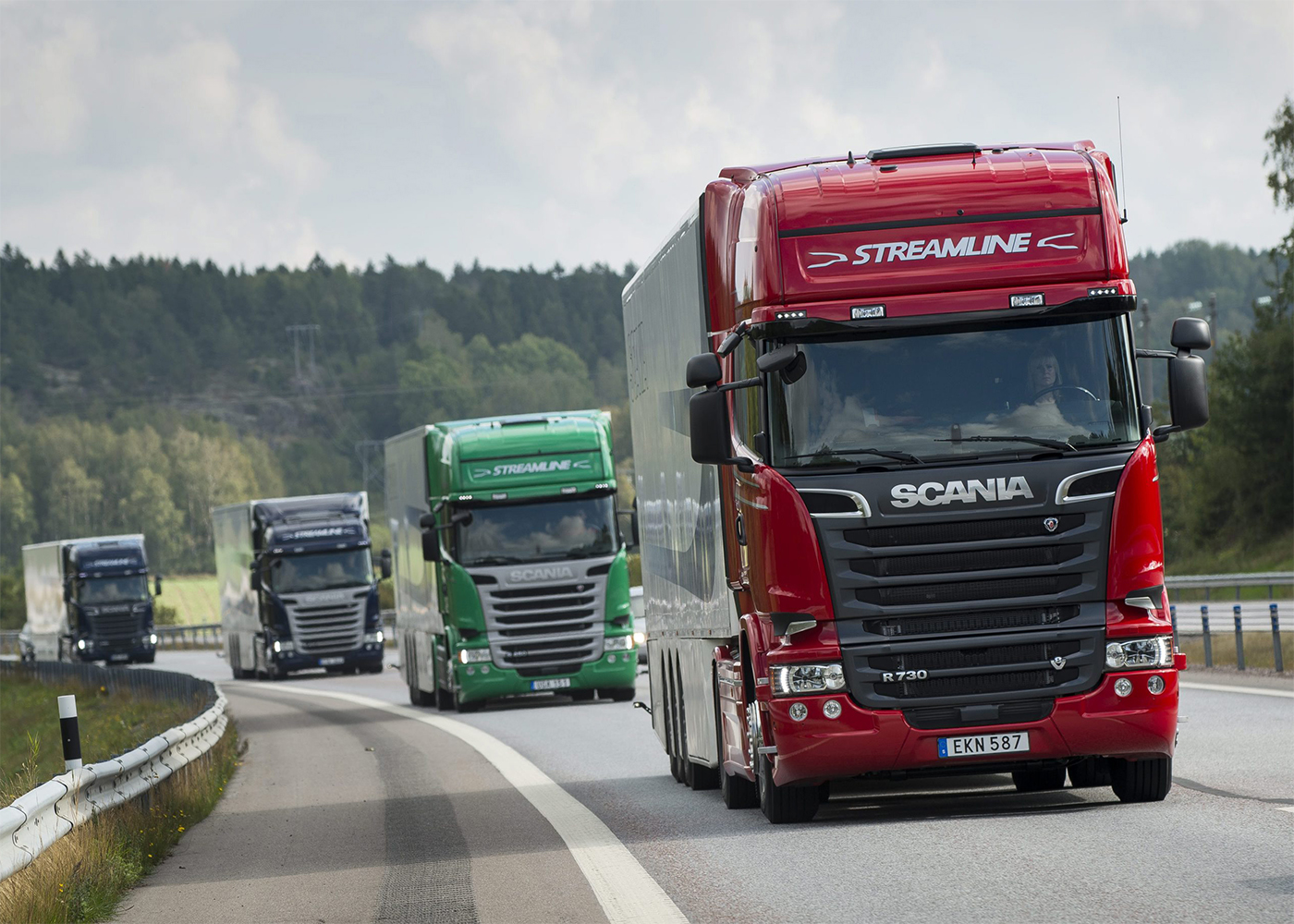}
  \vskip-1mm%
  \caption{Four heavy-duty vehicles in a platoon.}
  \label{fig_platoonphoto}
\end{center}
\end{figure}

Modern information and communication technologies enable such development, as the use of vehicle-to-vehicle and vehicle-to-infrastructure communication and the availability of ubiquitous computing power allow for the real-time coordination and automatic control of large groups of vehicles. In particular, the
formation of groups of closely-spaced heavy-duty vehicles allows them to cooperatively reduce fuel consumption through a reduction in aerodynamic drag, see Figure~\ref{fig_platoonphoto}. Experiments have shown that these platoons can lead to fuel savings of about 10\%~\cite{bonnet_2000,alam_2010}. Consequently, a cooperative approach offers great potential for developing a more efficient road freight transportation sector, especially since road transportation is currently mainly done by individual long-haulage vehicles that do not exploit the benefits of platoon formation.

\begin{figure}
\begin{center}
  \includegraphics[width=0.95\columnwidth,bb=0 0 1400 1000]{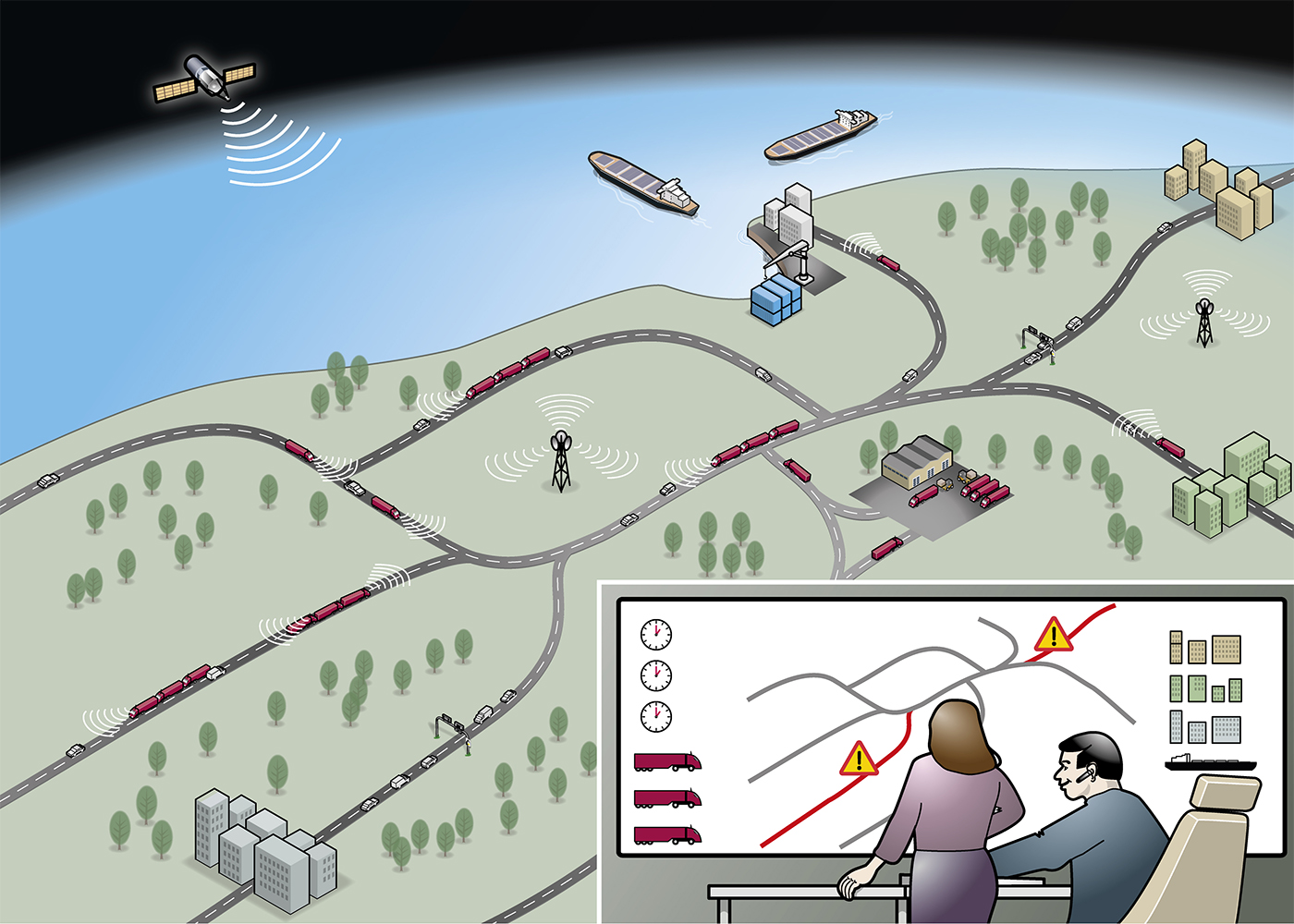}
  \vskip-1mm%
  \caption{An illustration of a cyber-physical approach to road freight transport with large-scale optimization of vehicle fleets and platoons.}
  \label{fig_cartoon_highlevel}
\end{center}
\end{figure}

In this paper, we present a cyber-physical approach to the control and coordination of a large fleet of heavy-duty vehicles that exploits the benefits of platooning, see Figure~\ref{fig_cartoon_highlevel}. In particular, a three-layer architecture is proposed that supports a hierarchical approach towards the minimization of the total fuel consumption. The bottom layer in this architecture deals with the automatic control of individual heavy-duty vehicles and is referred to as the vehicle layer. This vehicle control exploits vehicle-to-vehicle communications and advanced sensor technology to achieve a stable and safe platoon formation, leading to a reduction in fuel consumption through reduced aerodynamic drag. The middle layer, referred to as the cooperation layer, achieves additional fuel savings through the computation of fuel-optimal vehicle trajectories for the entire platoon. In addition, the formation of platoons is addressed in this layer through local decision-making and the execution of merging maneuvers. Finally, the fleet layer (i.e., the top layer) is aimed at the coordination of a potentially large fleet of vehicles belonging to multiple fleet owners. Here, the minimization of the fuel consumption is pursued by updating the plans of individual vehicles in order to achieve the most suitable platoon configurations.

Through the careful layering of the architecture, it is possible to significantly optimize the overall system performance while keeping the complexity at a manageable level. The tight integration of system components through vehicle-to-vehicle and vehicle-to-infrastructure communications as well as advanced on-board computations linked to cloud computations makes the road freight transportation system an excellent example of how major progress for such infrastructure applications is possible largely thanks to recent developments in cyber-physical systems.

The remainder of this paper is outlined as follows. Section~\ref{sec_opportunities} discusses the opportunities cyber-physical systems bring to freight transport and automated driving. Enabling information and communication technologies are reviewed and the proposed freight transport architecture is introduced. From the extensive amount of related literature on intelligent transportation systems and vehicle platooning, a small subset of the most relevant work is being treated in this section as well. The three layers of the freight transport architecture are presented in the next three sections. Section~\ref{sec_vehiclelayer} introduces the vehicle layer including the heavy-duty vehicle model, the vehicle control architecture, and the control strategy for vehicle platooning. The topics of look-ahead control for platooning and the coordination and control of merging maneuvers aim at optimizing the cooperative behavior of vehicle platoons and are treated in the cooperation layer discussed in Section~\ref{sec_cooperationlayer}. The fleet management layer is presented in Section~\ref{sec_fleetmanagementlayer} and coordinates the platoon planning and execution. Section~\ref{sec_evaluation} presents an evaluation of the freight transportation system through of a case study and is followed by the conclusions in Section~\ref{sec_conclusions}.

\section{Cyber-physical Systems Opportunities}\label{sec_opportunities}

\subsection{Enabling technologies}\label{sec_enablingtech}
Tremendous advances in computing, communication, and sensor technologies have enabled the current rapid development of intelligent transport systems \cite{gharavi_2007}. Today's high-end road vehicles are typically equipped with extensive computing capabilities, multiple radio interfaces, and radar, camera and other sensor devices. Low-cost wireless local and wide area network transceivers facilitate vehicle-to-vehicle and vehicle-to-infrastructure communications~\cite{hartenstein_2008,karagiannis_2011}. By integrating vehicular communication with existing sensor technologies, applications that enhance safety, efficiency, and driver comfort are being developed.

Another set of technologies that support cooperative transportation systems is given by cloud computing and service architectures~\cite{armbrust_2010}. They
offer large computing and storage capabilities together with a seamless integration of a diverse group of third-party tools and services. For vehicular and transportation applications, new possibilities are emerging to build systems spanning over large geographic areas with close to real-time data gathering and decision making~\cite{whaiduzzaman_2014}. Vehicular position and velocity data are an important example of such data that are readily available through various sensing devices including mobile phones~\cite{herrera+10}. Such data have proven to be very useful in many contexts including the understanding of road usage patterns in urban areas~\cite{wang+12}. For freight transportation it is shown in this paper how traffic, weather and other public and private data can be utilized in a transport planning and logistics application implemented through cloud technologies. The overall functional architecture of such a system is
described next.

\subsection{Freight transport system architecture}\label{sec_freighttransport}

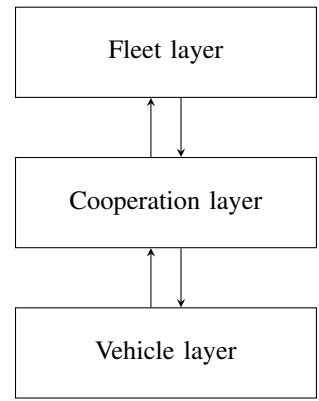
\begin{figure}
\begin{center}
\begin{tikzpicture}[
layer/.style={rectangle, minimum height=12mm, minimum width=40mm, draw=black}]
\def\dx{2mm};
\node (layer1) at (0,4) [layer] {Fleet layer};
\node (layer2) at (0,2) [layer] {Cooperation layer};
\node (layer3) at (0,0) [layer] {Vehicle layer};
\draw[-stealth] ([xshift= \dx]layer1.south) to ([xshift= \dx]layer2.north);
\draw[-stealth] ([xshift= \dx]layer2.south) to ([xshift= \dx]layer3.north);
\draw[-stealth] ([xshift=-\dx]layer2.north) to ([xshift=-\dx]layer1.south);
\draw[-stealth] ([xshift=-\dx]layer3.north) to ([xshift=-\dx]layer2.south);
\end{tikzpicture}
\caption{Layered freight transport system architecture.}
\label{fig_architecture}
\end{center}
\end{figure}

The freight transportation system discussed in this paper integrates potentially thousands of heavy-duty vehicles into a large-scale planning, cooperation, and real-time optimization and automation system. It is truly a complex and large-scale system built upon existing and emerging communication and computing infrastructures into a tightly coupled cyber-physical system with many human and social components. In order to manage the complexity of this large-scale coordination problem, the layered architecture in Figure~\ref{fig_architecture} is naturally adopted. Herein, the control of individual vehicles is addressed in the vehicle layer, whereas the cooperation layer targets the behavior and formation of platoons of vehicles. The large-scale coordination of vehicle fleets is handled in the fleet layer.

Specifically, the vehicle layer builds upon existing vehicle control systems to achieve the desired longitudinal behavior as needed to safely and automatically operate vehicles and vehicle platoons. Hereto, a decentralized controller is synthesized that exploits vehicle-to-vehicle communication and advanced sensor
information (e.g., radar) to guarantee the tracking of a specified inter-vehicular distance as well as the rejection of disturbances. We recall that the operation of vehicles in closely-spaced platoons reduces fuel consumption.

\begin{figure}
\begin{center}
  \includegraphics[width=0.95\columnwidth,bb=0 0 1400 1000]{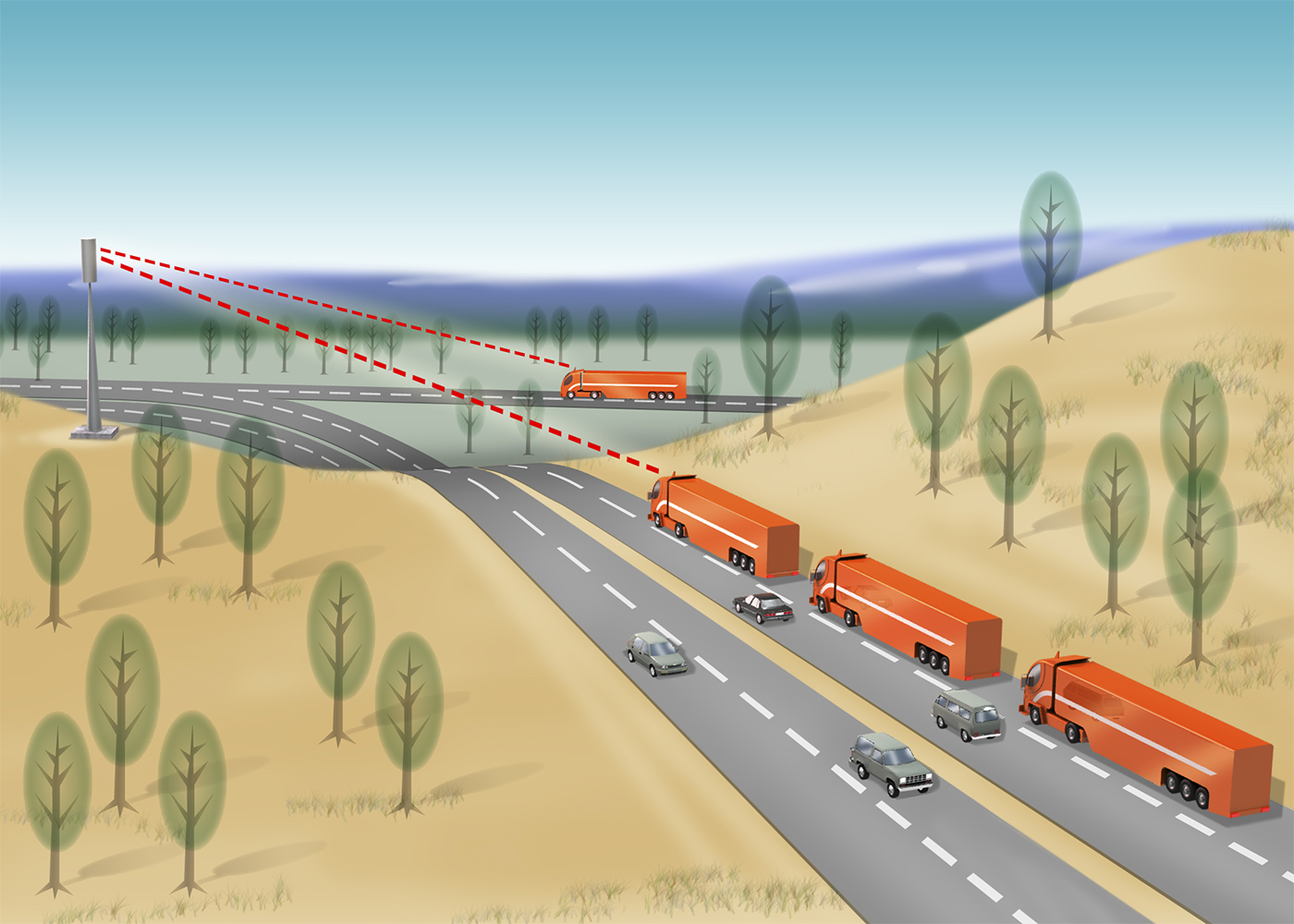}
  \vskip-1mm%
  \caption{Automatic control of a merging maneuver, where a single vehicle is about to merge with a three-vehicle platoon after a road intersection. The operation is supported by vehicle-to-infrastructure communication.}
  \label{fig_cartoon_lowlevel}
\end{center}
\end{figure}

The aim of the cooperation layer is twofold. First, it computes fuel-optimal velocity profiles for vehicle platoons taking road topography and traffic into account. For example, by exploiting look-ahead information about the road topography when driving over hilly terrain, braking can be avoided and additional fuel savings can be obtained. Second, the cooperation layer locally coordinates the behavior of vehicles or platoons with overlapping route segments by deciding whether neighbouring vehicles should form a platoon. In addition to this decision-making process, the optimal control of merging maneuvers for platoon formation is handled in this layer, as illustrated in Figure~\ref{fig_cartoon_lowlevel}. Vehicle-to-vehicle and vehicle-to-infrastructure communication are exploited for this coordination, which extend only to the relative vicinity of the vehicle and platoon.

The fleet layer targets the large-scale coordination over a significant geographic area for a large group of vehicles from potentially different fleet owners, see Figure~\ref{fig_cartoon_highlevel}. By updating the routes and transport plans of individual vehicles, the formation of platoons can be encouraged and the total fuel consumption of the fleet can be minimized. In addition to this coordination, the fleet layer includes the task of transport planning to target a better utilization of the capacity of the freight transport system. Optimization criteria in this layer can incorporate not only costs directly associate with individual fleet owners, but can include societal aspects such as traffic congestion and environment impact.

The layers in Figure~\ref{fig_architecture} are presented in some more detail in Sections~\ref{sec_vehiclelayer} to~\ref{sec_fleetmanagementlayer} and the overall system is evaluated in a case study in Section~\ref{sec_evaluation}, but first we give a brief overview of related work.

\subsection{Background on vehicle platooning}\label{sec_relatedwork}
The freight transport architecture in Figure~\ref{fig_architecture} is motivated by the concept of an automated highway system~\cite{varaiya_1993,horowitz_2000}, in which cars are organized in platoons to increase traffic flow. Further examples of such systems are given in~\cite{raza_1996} and~\cite{tsugawa_2000}. The layers in these architectures typically range from vehicles in the bottom layer to a road network in the top layer. Our architecture focuses on heavy-duty vehicles and aims at optimizing the transportation of goods rather than traffic flow in general. We note that similar architectures are also used in many related engineering systems, such as air traffic management~\cite{zhang_2012} and spacecraft formation~\cite{beard_2001}.

The idea of highway automation and platooning has a long history, with first visions dating back at least to the 1930s~\cite{gm_film_1940}. Apart from early analysis of the dynamics of vehicle following~\cite{chandler_1958}, the first control strategies for vehicle platooning appeared in~\cite{levine_1966}
and~\cite{melzer_1971,chu_1974}. Many results have appeared since, focusing on topics ranging from analysis of spacing policies~\cite{ioannou_1993,swaroop_1994} to experimental validation~\cite{naus_2010}. For heavy-duty vehicles, platooning is mainly  motivated by a reduced fuel consumption and several experimental
evaluations have focussed on this aspect~\cite{bonnet_2000,lammert_2014,alam_2015b}.

The operation of platoons relies on the (partial) automation of heavy-duty vehicles. Large research efforts are being undertaken in the development of fully autonomous vehicles, of which an early prototype is discussed in~\cite{dickmanns_1994}. Several challenges organized by DARPA have further spurred development in this area~\cite{buehler_2008}, whereas a recent overview is given in~\cite{bengler_2014}.

\section{Vehicle layer}\label{sec_vehiclelayer}

\subsection{Vehicle model}\label{sec_vehiclemodel}
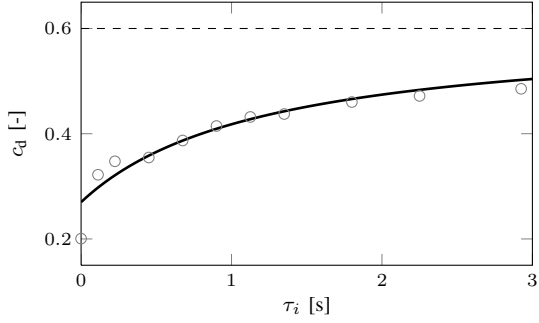
\begin{figure}
\begin{center}
\begin{tikzpicture}
\begin{axis}[
  height=35mm,
  width=60mm,
  at={(0,0)}, scale only axis,
  xlabel={$\tau_i$ [s]},
  ylabel={$c_{\text{\rm d}}$ [-]},
  xmin=0, xmax=3, ymin=0.15, ymax=0.65,
  x label style={font=\footnotesize, yshift=1mm},
  y label style={font=\footnotesize},
  tick label style={font=\scriptsize},
]
\addplot[gray,mark=o,only marks] table {img/airdragdata_experiment.dat};
\addplot[black,line width=1pt] table {img/airdragdata_function.dat};
\addplot[black,dashed] coordinates { (0,0.6) (3,0.6) };
\end{axis}
\end{tikzpicture}
\vskip-2mm%
\caption{Air drag coefficient $c_{\rm d}(\tau_i)$ as a function of time gap $\tau_i$ for $c_{\rm d}^0 = 0.6$, $\alpha_1 = 0.53$, and $\alpha_2 = 0.81$. The function is estimated based on experimental data (circles) reported in~\cite{book_hucho_1998}.}
  \label{fig_airdragcoefficient}
\end{center}
\end{figure}

The heavy-duty vehicle control and cooperation algorithms are based on a dynamic model of the powertrain. Specifically, the longitudinal dynamics of a vehicle indexed~$i$ is modeled~as
\begin{align}
\begin{split}
\dot{s}_i &= v_i, \\
\!m\dot{v}_i &= -F_{\text{\rm r}}(\alpha(s_i)) - F_{\text{\rm g}}(\alpha(s_i)) - F_{\text{\rm d}}(\dt_i,v_i) \\ &\phantom{=}\qquad + F_{\text{\rm e},i} - F_{\text{\rm b},i}.
\end{split}\label{eqn_vehiclemodel}
\end{align}
Here, $s_i$ and $v_i$ denote its longitudinal position and velocity, respectively, which are collected in the state $x_i =(s_i,v_i)^{\T}$. For ease of exposition, we let all vehicles have identical parameter values, but the results in the paper extend directly to heterogeneous vehicle groups. In (\ref{eqn_vehiclemodel}), $m$ represents the vehicle mass, whereas $F_{\text{\rm r}}$ and $F_{\text{\rm g}}$ denote the rolling resistance and the longitudinal component of gravity, respectively. The latter is given as
\begin{align}
F_{\text{\rm g}}(\alpha(s_i)) &= mg\sin\alpha(s_i),
\label{eqn_forces_gravity}
\end{align}
with $\alpha(s_i)$ the road gradient at position $s_i$ and $g$ the gravitational acceleration. The aerodynamic drag $F_{\text{\rm d}}$ satisfies
\begin{align}
F_{\text{\rm d}}(\dt_i,v_i) = \tfrac{1}{2} c_{\text{\rm d}}(\dt_i) \rho A v_i^2,
\label{eqn_forces_airdrag}
\end{align}
where $\rho$ is the air density and $A$ denotes the frontal area of the vehicle. The air drag is dependent on the time gap $\dt_i$ between vehicle $i$ and its predecessor, as captured through the air drag coefficient $c_{\text{\rm d}}(\dt_i)$. Here, the time gap represents the time difference between two successive vehicle passing the same point on the road. The air drag coefficient is modeled as
\begin{align}
c_{\text{\rm d}}(\dt_i) = c_{\text{\rm d}}^0 \left(1 - \frac{\alpha_1}{1 + \alpha_2\dt_i}\right),
\label{eqn_airdragcoefficient}
\end{align}
where $c_{\text{\rm d}}^0$ represents the nominal air drag coefficient for a heavy-duty vehicle driving alone, and the parameters $\alpha_1$ and $\alpha_2$ characterize the air drag reduction as the time gap between vehicles decreases. Figure~\ref{fig_airdragcoefficient} shows an illustration of $c_{\text{\rm d}}(\dt_i)$ as estimated from experimental data. This air drag reduction obtained through smaller inter-vehicular distances offers a potential for saving fuel, which is extensively exploited throughout the paper.

Finally, the forces $F_{\text{\rm e},i}$ and $F_{\text{\rm b},i}$ in (\ref{eqn_vehiclemodel}) denote the traction force at the wheels and the force exerted by the brakes, respectively. They are control inputs. The corresponding injected fuel flow $\varphi_i$ depends on the instantaneous power $F_{\text{\rm e},i}v_i$, which is bounded as $P_{\text{\rm min}} \leq F_{\text{\rm e},i}v_i \leq P_{\text{\rm max}}$, and obtained through
\begin{align}
\varphi_i = p_1F_{\text{\rm e},i}v_i + p_0.
\label{eqn_fuelflow}
\end{align}
Here, the parameters $p_0$ and $p_1$ aggregate the effects of engine and gear box efficiency. Specifically, $p_0$ captures the fuel flow when the engine is idling. Consequently, the nominal fuel consumption (normalized with respect to the travelled distance) of a single vehicle reads
\begin{align}
J_{\text{\rm nom},i} = \frac{1}{s_i(t^0) - s_i(t^1)}\int_{t^0}^{t^1} \varphi_i(t) \di t,
\end{align}
for any time interval satisfying $t^1-t^0>0$. The remainder of this paper will be focused on systematically reducing the nominal fuel consumption by exploiting platooning.

\subsection{Vehicle control architecture}\label{sec_vehiclecontrolarchitecture}

\begin{figure}
\begin{center}
\begin{tikzpicture}[
  box/.style={rectangle, draw, align=center, inner sep=0mm, minimum width=13mm, minimum height=6mm},
  control/.style={rectangle, draw, align=center, inner sep=0mm, minimum width=20mm, minimum height=10mm},
  data/.style={rectangle, draw, align=center, inner sep=0mm, minimum width=20mm, minimum height=10mm},
  canbus/.style={},
  signal/.style={semithick, >=stealth},
]
\node (radar) at (-32mm,9mm) [box] {Radar};
\node (gps) at (-32mm,0mm) [box] {\textsc{gps}};
\node (wsu) at (-32mm,-9mm) [box] {\textsc{wsu}};
\node (ems) at (32mm,9mm) [box] {\textsc{ems}};
\node (bms) at (32mm,0mm) [box] {\textsc{bms}};
\node (gms) at (32mm,-9mm) [box] {\textsc{gms}};
\node (can) at (0mm,15mm) [] {\textsc{can} bus};
\draw[canbus] (-19mm,-12mm) to (-19mm,14mm) to (-8mm,14mm);
\draw[canbus] (-20mm,-12mm) to (-20mm,15mm) to (-8mm,15mm);
\draw[canbus] (-21mm,-12mm) to (-21mm,16mm) to (-8mm,16mm);
\draw[canbus] (19mm,-12mm) to (19mm,14mm) to (8mm,14mm);
\draw[canbus] (20mm,-12mm) to (20mm,15mm) to (8mm,15mm);
\draw[canbus] (21mm,-12mm) to (21mm,16mm) to (8mm,16mm);
\draw[signal,->] (radar.east) to (-21mm,9mm);
\draw[signal,->] (gps.east) to (-21mm,0mm);
\draw[signal,->] (wsu.east) to (-21mm,-9mm);
\draw[signal,<-] (ems.west) to (21mm,9mm);
\draw[signal,<-] (bms.west) to (21mm,0mm);
\draw[signal,<-] (gms.west) to (21mm,-9mm);
\node (control) at (0mm,-7mm) [control] {Vehicle\\ controller};
\node (data) at (0mm,6mm) [data] {Data\\ processing};
\draw[signal,->] (data.south) to (control.north);
\draw[signal,<-] (data.west) to (-19mm,6mm);
\draw[signal,->] (control.east) to (19mm,-7mm);
\end{tikzpicture}
\caption{A controller area network enables the communication of sensor
  data to the vehicle controller, which computes control commands to
  be executed by the engine, braking, and gear management systems.}
\label{fig_vehiclecontrolarchitecture}
\end{center}
\end{figure}
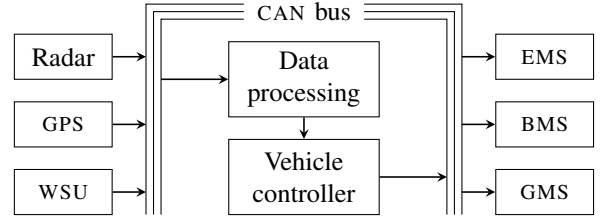
The vehicle control architecture for the powertrain is depicted in Figure~\ref{fig_vehiclecontrolarchitecture}. A controller area network~\cite{johansson_2005} inside the vehicle communicates radar and positioning data together with data from other vehicles through the wireless sensor unit to a data processing unit. The vehicle controller computes low-level commands and sends them to the engine management system, brake management system, and gear management system. These systems implement the desired longitudinal vehicle behavior. Automatic velocity control is often achieved by letting the vehicle controller execute cruise controller or adaptive cruise controller algorithms. A cruise controller uses measurements of the vehicle speed to maintain a constant reference velocity in order to improve fuel economy and driver comfort. The adaptive cruise controller is an extension that includes radar information to obtain an estimate of the position and velocity of a preceding vehicle, improving safety and convenience.

Next, an alternative vehicle controller is presented that exploits additional information about the preceding vehicle obtained through wireless communication. By sharing information, automatic control of small inter-vehicular distances with guaranteed safety can be achieved.

\subsection{Vehicle control for platooning}\label{sec_platooncontrol}

\begin{figure}
\begin{center}
\begin{tikzpicture}[
  vehicle/.style={rectangle, draw, align=center, inner sep=0mm, minimum height=17mm, minimum width=30mm},
  separator/.style={thin,dashed},
  control/.style={rectangle, draw, align=center, inner sep=0mm, minimum height=12mm, minimum width=30mm},
  localcontrol/.style={rectangle, align=center, inner sep=0mm, minimum height=4mm, minimum width=10mm},
  clac/.style={rectangle, draw, align=center, inner sep=0mm, minimum height=12mm, minimum width=70mm},
  signal/.style={semithick,>=stealth},
]
\def\dx{20mm}
\node (v1) at (-\dx,-2.5mm) [vehicle] {};
\node (scania1) at ([yshift=6mm]v1.south) {\includegraphics[width=24mm]{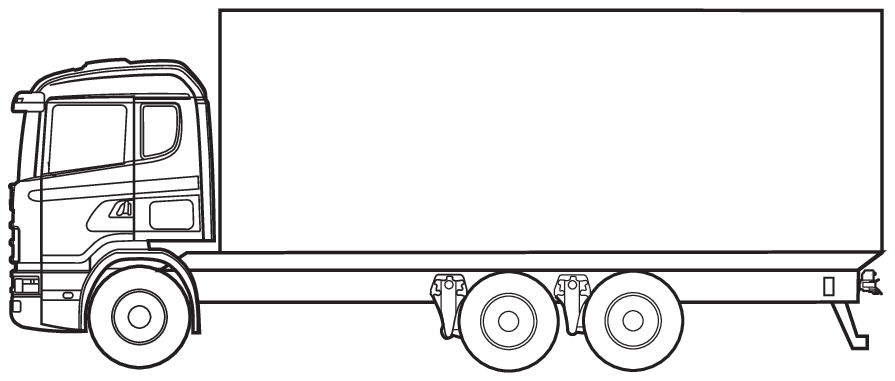}};
\node (bms1) at ([yshift=-2.5mm]v1.north) [localcontrol] {\textsc{bms}};
\node (ems1) at ([xshift=-10mm]bms1) [localcontrol] {\textsc{ems}};
\node (gms1) at ([xshift= 10mm]bms1) [localcontrol] {\textsc{gms}};
\draw[separator] (bms1.north west) to (bms1.south west);
\draw[separator] (bms1.north east) to (bms1.south east);
\draw[separator] (ems1.south west) to (gms1.south east);
\node (control1) at (-\dx,20mm) [control] {Vehicle $i-1$\\ platoon controller};
\draw[->,signal] ([xshift=-2mm]v1.north) to ([xshift=-2mm]control1.south);
\draw[<-,signal] ([xshift= 2mm]v1.north) to ([xshift= 2mm]control1.south);
\node (v2) at (\dx,-2.5mm) [vehicle] {};
\node (scania2) at ([yshift=6mm]v2.south) {\includegraphics[width=24mm]{img/scania_small.eps}};
\node (bms2) at ([yshift=-2.5mm]v2.north) [localcontrol] {\textsc{bms}};
\node (ems2) at ([xshift=-10mm]bms2) [localcontrol] {\textsc{ems}};
\node (gms2) at ([xshift= 10mm]bms2) [localcontrol] {\textsc{gms}};
\draw[separator] (bms2.north west) to (bms2.south west);
\draw[separator] (bms2.north east) to (bms2.south east);
\draw[separator] (ems2.south west) to (gms2.south east);
\node (control2) at (\dx,20mm) [control] {Vehicle $i$\\ platoon controller};
\draw[->,signal] ([xshift=-2mm]v2.north) to ([xshift=-2mm]control2.south);
\draw[<-,signal] ([xshift= 2mm]v2.north) to ([xshift= 2mm]control2.south);
\node (clac) at (0,40mm) [clac] {Cooperative look-ahead control};
\draw[->,signal] ([xshift=-2mm]control1.north) to ([xshift=-2mm]clac.south-|control1.north);
\draw[<-,signal] ([xshift= 2mm]control1.north) to ([xshift= 2mm]clac.south-|control1.north);
\draw[->,signal] ([xshift=-2mm]control2.north) to ([xshift=-2mm]clac.south-|control2.north);
\draw[<-,signal] ([xshift= 2mm]control2.north) to ([xshift= 2mm]clac.south-|control2.north);
\draw[->,signal] (control1.east) to (control2.west);
\draw[-,separator] (-37mm-6mm,30mm) to (37mm,30mm);
\node at (-41mm,7.5mm) [align=center, rotate=90] {Vehicle layer};
\draw[-,separator] (-37mm-6mm,-15mm) to (37mm,-15mm);
\end{tikzpicture}
\caption{Control architecture corresponding to the vehicle layer in Figure~\ref{fig_architecture}.}
\label{fig_architecture_lowerlayers}
\end{center}
\end{figure}

This section presents a strategy for the longitudinal control of heavy-duty vehicle platoons. It is positioned in the vehicle layer of the freight transport architecture in Figure~\ref{fig_architecture} and is detailed in Figure~\ref{fig_architecture_lowerlayers}. The objective of the platoon controller is to achieve small inter-vehicular distances while tracking a varying reference velocity $\vref(\cdot)$. The reference velocity, which is specified as a function of the position on the road, is the result of the cooperative look-ahead control strategy that is discussed in Section~\ref{sec_clac}.

As it is well-known that standard policies for specifying the inter-vehicular distance in a platoon are not compatible with tracking a spatially varying reference velocity profile~\cite{alam_2013,besselink_2015b}, we adopt the delay-based spacing policy
\begin{align}
s_{\text{\rm ref},i}(t) = s_{i-1}(t-\dtref),
\label{eqn_spacingpolicy_delaybased}
\end{align}
where $s_{\text{\rm ref},i}$ denotes the desired longitudinal position of vehicle $i$. It is convenient to express (\ref{eqn_spacingpolicy_delaybased}) in the spatial domain. To this end, let the spatial position $s$ be the independent variable and denote $t_i(s)$ as the time instance at which vehicle $i$ passes $s$. By introducing the time gap tracking errors
\begin{align}
\Delta_i(s) &= t_i(s) - t_{i-1}(s) - \dtref, \label{eqn_spacingpolicy_Deltai}\\
\Delta_i^0(s) &= t_i(s) - t_0(s) - i\dtref, \label{eqn_spacingpolicy_Deltai0}
\end{align}
the policy (\ref{eqn_spacingpolicy_delaybased}) is equivalent to $\Delta_i = 0$. The condition (\ref{eqn_spacingpolicy_Deltai0}) represents the time gap tracking error with respect to the first vehicle in a platoon. Similarly, a velocity tracking error $e_i$ can be defined for each vehicle, representing the deviation from the desired reference velocity profile $\vref(s)$. On the basis of the time gap and velocity tracking errors, a weighted error signal is introduced as
\begin{align}
\delta_i(s) = (1-h_0)\Delta_i(s) + h_0\Delta_i^0(s) + he_i(s),
\label{eqn_spacingpolicy_deltai}
\end{align}
in which the parameters $0\leq h_0 < 1$ and $h>0$ provide a measure of the influence of the lead vehicle and velocity tracking, respectively.

A distributed controller design can be achieved on the basis of the weighted error signal (\ref{eqn_spacingpolicy_deltai}) and powertrain dynamics (\ref{eqn_vehiclemodel}), hereby satisfying two objectives. First, the controller for vehicle $i$ should guarantee the existence of a unique equilibrium point for which $\delta_i = 0$ and, second, it should asymptotically stabilize this equilibrium. Namely, any controller that achieves this ensures asymptotic stability of the desired spacing policy (\ref{eqn_spacingpolicy_delaybased}) throughout the platoon. A controller based on feedback linearization that achieves these objectives is given in \cite{besselink_2015b}. It is stressed that, as each vehicle individually addresses the local goal of achieving $\delta_i\rightarrow0$, the controller is distributed. Herein, vehicles exploit radar measurements as well as information from the preceding vehicle and (potentially) the lead vehicle obtained through wireless communication.

For any controller that asymptotically stabilizes the equilibrium corresponding to $\delta_i = 0$, it can be shown that the velocity tracking errors of two successive vehicles satisfy
\begin{align}
\int_0^s |e_i(\sigma)|^2 \di\sigma \leq \int_0^s |e_{i-1}(\sigma)|^2 \di\sigma,
\label{eqn_stringstability}
\end{align}
which indicates that any perturbations do not grow as they propagate through the platoon. The inequality is strict when information of the lead vehicle is included, i.e., $h_0>0$, which also ensures robustness with respect to external disturbances acting on the vehicles, see \cite{besselink_2015b} and related work in \cite{seiler_2004}. Condition (\ref{eqn_stringstability}) is an example of string stability, which provides stability notions for vehicle platoons. An early notion of string stability can be found in~\cite{peppard_1974}, whereas a formal definition is given in~\cite{swaroop_1996}. For an overview and examples of alternative definitions, see \cite{ploeg_2014} and \cite{fenton_1968,sheikholeslam_1993}, respectively.

\section{Cooperation layer}\label{sec_cooperationlayer}

\subsection{Cooperative look-ahead control}\label{sec_clac}
The aim of the cooperative look-ahead control strategy is to compute a velocity profile $\vref(\cdot)$ that is feasible for each individual heavy-duty vehicle in the platoon and fuel-optimal for the overall platoon. The speed profile is communicated to the vehicle layer, as described in Figure~\ref{fig_architecture_lowerlayers}, where each vehicle controller tracks $\vref(\cdot)$ while guaranteeing stability and safety. The computation of the speed profile is accomplished by solving a receding horizon control problem that includes the dynamics and corresponding constraints of each vehicle and minimizes a cost function depending on the fuel consumption of the whole platoon. To this end, the vehicle model (\ref{eqn_vehiclemodel}) is expressed in the spatial domain as is the constraint that all vehicles in the platoon track the same velocity profile:
\begin{equation}
v_i(s)=v_\text{ref}(s), \quad i=1,\ldots,N,
\label{eq:clac_same_speed}
\end{equation}
where $N$ is the number of vehicles in the platoon. Note that the delay-based spacing policy (\ref{eqn_spacingpolicy_delaybased}) also requires equal velocity profiles in the spatial domain, which therefore corresponds to the constraint (\ref{eq:clac_same_speed}). Moreover, as the road altitude is dependent on the position, this policy is well-suited for platooning over road segments with varying topography~\cite{turri_2015}.

The cost function for the cooperative look-ahead controller to minimize is defined as the sum of the fuel consumption for all vehicles in the platoon:
\begin{equation}
J_{\textsc{clac}} = \frac{1}{NH} \sum_{i=1}^N \int_{s^0}^{s^0 + H} \varphi_i(s) \frac{1}{\vref(s)} \di s,
\label{eq:clac_cost_function}
\end{equation}
where $\varphi_i$ is the injected fuel flow (\ref{eqn_fuelflow}) (expressed in the spatial domain), $s_0$ the current position of the leading vehicle (i.e., $s_1(t)$) and $H$ the horizon length. The average speed request for a given road segment as imposed by the fleet management layer is denoted by $\bar{v}$ and is enforced through the constraint
\begin{equation}
\frac{1}{H}\int_{s^0}^{s^0+H} \frac{1}{v_\text{ref}(s)} \di s = \frac{1}{\bar{v}}.
\label{eq:clac_constraint_average_speed}
\end{equation}

The cooperative look-ahead controller is implemented with a receding horizon and can be summarized as follows:
\begin{equation}
\begin{aligned}
\min \quad & \text{Platoon fuel consumption \eqref{eq:clac_cost_function}} \\
\text{subj. to} \quad
& \text{Vehicle dynamics \eqref{eqn_vehiclemodel} (in the spatial domain),} \\
& \text{Constraints on state and input,} \\
& \text{Constraint on the average velocity \eqref{eq:clac_constraint_average_speed},} \\
& \text{Common platoon velocity \eqref{eq:clac_same_speed}}.
\end{aligned}
\nonumber
\end{equation}
Here, the constraints on state and input refer to the speed limits as well as the bounds on engine power and braking force. The receding horizon problem can be solved using dynamic programming~\cite{book_bellman_1957}, see~\cite{turri_2015} for details. In the special case that the platoon consists of only $N=1$ vehicle, the proposed platoon controller corresponds to the single-vehicle look-ahead controller~\cite{hellstrom_2009}.

Altitude variations have a significant impact on the behavior of heavy-duty vehicles. Due to their inertia and limited engine power, they are typically not able to maintain a constant velocity while driving over steep up-slopes and down-slopes. This effect is critical when a group of vehicles, that can significantly differ in mass and powertrain characteristic, needs to maintain the short inter-vehicular distances required by platooning. Experimental results have for instance shown how follower vehicles in a platoon driving downhill need to brake in order to compensate their different inertia and experienced air drag force~\cite{alam_2015b}. Therefore, the particular structure of the cooperative look-ahead controller proposed here with common velocity profiles~(\ref{eq:clac_same_speed}) seems to have several advantages~\cite{turri_2015}. Earlier work on look-ahead control for the fuel-efficient traversal of hilly road segments has focussed on single vehicles only, with early work considering simple road profiles and exploiting analytical solutions~\cite{schwarzkopf_1977,stoicescu1995fuel}. Algorithms based on dynamic programming suitable for more generic road profiles have also been proposed~\cite{hooker1988optimal,monastyrsky1993rapid,hellstrom_2009}.

\subsection{Optimal control of merging maneuvers}\label{sec_merging}
Let us now focus on the formation of platoons through the merging of individual vehicles or platoons that approach a common point after a highway intersection or an on-ramp. This maneuver is essential for platoon formation and it enables the high-level coordination of platoons as will be discussed in Section~\ref{sec_opportunisticplatoonformation}.

\begin{figure}
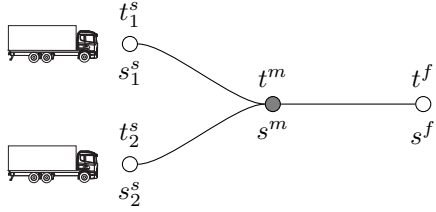

\begin{center}
\begin{tikzpicture}[
  startingpoint/.style={draw,circle,minimum size=2mm,inner sep=0mm},
  mergingpoint/.style={draw,circle,minimum size=2mm,inner sep=0mm,fill=gray},
  finalpoint/.style={draw,circle,minimum size=2mm,inner sep=0mm},
  edge/.style={},
]
\node[xscale=-1,anchor=west] (scania1) at (-22mm,8mm) {\includegraphics[width=12mm]{img/scania_small.eps}};
\node[xscale=-1,anchor=west] (scania2) at (-22mm,-8mm) {\includegraphics[width=12mm]{img/scania_small.eps}};
\node (s1) at ([xshift=3mm]scania1.west) [startingpoint,label={below:$s_1^s$},label={above:$t_1^s$}] {};
\node (s2) at ([xshift=3mm]scania2.west) [startingpoint,label={below:$s_2^s$},label={above:$t_2^s$}] {};
\node (m) at (0mm,0mm) [mergingpoint,label={above:$t^m$},label={below:$s^m$}] {};
\node (f) at (20mm,0mm) [finalpoint,label={above:$t^f$},label={below:$s^f$}] {};
\draw[edge] (m) to (f);
\draw[edge] (s1) .. controls ([xshift=5mm]s1) and ([xshift=-5mm]m) .. (m);
\draw[edge] (s2) .. controls ([xshift=5mm]s2) and ([xshift=-5mm]m) .. (m);
\end{tikzpicture}
\vskip-2mm%
\caption{Schematic illustration of a two-vehicle optimal merging problem.}
\label{fig_mergingproblem}
\end{center}
\end{figure}

Consider the simple merging problem for two vehicles $i=1,2$ illustrated in Figure~\ref{fig_mergingproblem}. Here,~$s^m$ denotes the location of an intersection and $s_1^s$, $s_2^s$ the positions on two road segments from which the merging maneuver is initiated. The times $t_1^s$ and $t_2^s$  at which the vehicles arrive at these positions are taken as the starting times for the merging maneuver, for which the initial states $x_i^s =(s_i^s,v_i^s)^{\T}$, $i=1,2$, hold for some
velocity~$v_i^s$. A common final state $x^f = (s^f,v^f)^{\T}$ and time $t^f$ is chosen after the intersection to obtain the desired average velocity over the road segment. Suppose the vehicles merge to form a platoon at $s^m$ at time $t^m$, so that approximately
\begin{align}
x_1(t) = x_2(t), \quad \forall t\in[t^m,t^f].
\label{eqn_merging_constraint}
\end{align}
The merging time $t^m$ is not fixed a priori, but is the result of an optimization. Due to a reduced aerodynamic drag, the vehicle dynamics and the total fuel cost is obviously different after the merging point compared to before. Therefore, the total fuel consumption for the overall operation can be expressed~as
\begin{align}
\sum_{i=1}^2 \int_{t_i^s}^{t^m} \varphi_i(t) \di t + \int_{t^m}^{t^f} \sum_{i=1}^2 \varphi_i(t) \di t.
\label{eqn_merging_cost_rewritten}
\end{align}
This cost function can be minimized subject to the dynamics (\ref{eqn_vehiclemodel}) using a two-step hybrid optimal control approach~\cite{sussmann_1999,shaikh_2007}, as detailed in~\cite{koller_2015}. In the first step, after selecting a fixed merging time $t^m$, the problem reduces to the fuel-optimal traversal of a given road segment. The partitioning of the total cost in (\ref{eqn_merging_cost_rewritten}) corresponds exactly to these road segments. For the last road segment traversed as a platoon, the platoon dynamics satisfy the constraint (\ref{eqn_merging_constraint}). In the second step, an optimization of the merging time $t^m$ is performed. When this process is repeated iteratively, the optimality of the overall problem can be guaranteed through the hybrid maximum principle~\cite{sussmann_1999}, which is an extension of the Pontryagin maximum principle. The two-vehicle merging problem discussed here is easily extended to cases in which the optimization includes more vehicles, constraints on the desired velocity at the merging instant, and successive merging
maneuvers. Moreover, a receding horizon implementation of the optimal merging procedure can be used to guarantee robustness with respect to disturbances such as the influence of surrounding traffic. These extensions can be found in~\cite{koller_2015}.

\subsection{Opportunistic platoon formation}\label{sec_opportunisticplatoonformation}

\begin{figure}
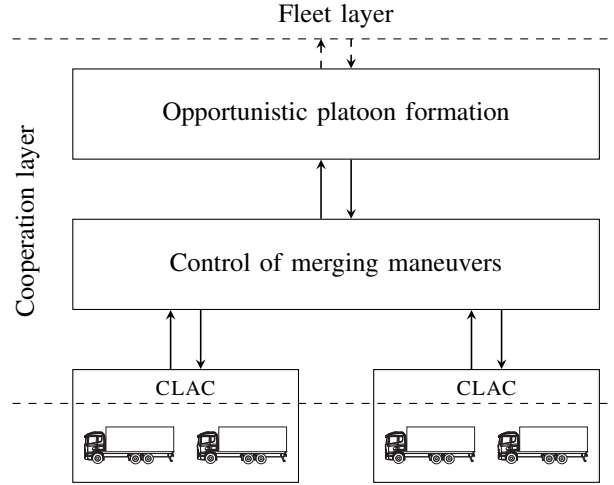

\begin{center}
\begin{tikzpicture}[
  platoon/.style={rectangle, draw, align=center, inner sep=0mm, minimum height=15mm, minimum width=30mm},
  separator/.style={thin,dashed},
  localcontrol/.style={rectangle, align=center, inner sep=0mm, minimum height=4mm, minimum width=30mm},
  merging/.style={rectangle, draw, align=center, inner sep=0mm, minimum height=12mm, minimum width=70mm},
  adhoc/.style={rectangle, draw, align=center, inner sep=0mm, minimum height=12mm, minimum width=70mm},
  signal/.style={semithick,>=stealth},
]
\def\dx{20mm}
\node (platoon1) at (-\dx,-1.5mm) [platoon] {};
\node (clac1) at ([yshift=-2.5mm]platoon1.north) [localcontrol] {\textsc{clac}};
\node (scania11) at ([xshift=-7.5mm,yshift=5mm]platoon1.south) {\includegraphics[width=12mm]{img/scania_small.eps}};
\node (scania12) at ([xshift=7.5mm,yshift=5mm]platoon1.south) {\includegraphics[width=12mm]{img/scania_small.eps}};
\node (platoon2) at (\dx,-1.5mm) [platoon] {};
\node (clac2) at ([yshift=-2.5mm]platoon2.north) [localcontrol] {\textsc{clac}};
\node (scania21) at ([xshift=-7.5mm,yshift=5mm]platoon2.south) {\includegraphics[width=12mm]{img/scania_small.eps}};
\node (scania22) at ([xshift=7.5mm,yshift=5mm]platoon2.south) {\includegraphics[width=12mm]{img/scania_small.eps}};
\node (merging) at (0,20mm) [merging] {Control of merging maneuvers};
\draw[->,signal] ([xshift=-2mm]platoon1.north) to ([xshift=-2mm]merging.south-|platoon1.north);
\draw[<-,signal] ([xshift= 2mm]platoon1.north) to ([xshift= 2mm]merging.south-|platoon1.north);
\draw[->,signal] ([xshift=-2mm]platoon2.north) to ([xshift=-2mm]merging.south-|platoon2.north);
\draw[<-,signal] ([xshift= 2mm]platoon2.north) to ([xshift= 2mm]merging.south-|platoon2.north);
\node (adhoc) at (0mm,40mm) [adhoc] {Opportunistic platoon formation};
\draw[<-,signal] ([xshift=-2mm]adhoc.south) to ([xshift=-2mm]merging.north-|adhoc.south);
\draw[->,signal] ([xshift= 2mm]adhoc.south) to ([xshift= 2mm]merging.north-|adhoc.south);
\draw[-,separator] (-37mm-6mm,1.5mm) to (37mm,1.5mm);
\draw[-,separator] (-37mm-6mm,50mm) to (37mm,50mm);
\node at (-41mm,25.75mm) [align=center, rotate=90] {Cooperation layer};
\node at (0mm,53mm) [align=center] {Fleet layer};
\draw[<-,signal,dashed] (-2mm,50mm) to ([xshift=-2mm]adhoc.north);
\draw[->,signal,dashed] (2mm,50mm) to ([xshift= 2mm]adhoc.north);
\end{tikzpicture}
\caption{Platoon coordination architecture according to the
  cooperation layer in Figure~\ref{fig_architecture}. The lower blocks
  correspond to the cooperative look-ahead control of the platoons
  linked to the vehicle layer in Figure~\ref{fig_architecture_lowerlayers}.}
\label{fig_architecture_upperlayers}
\end{center}
\end{figure}

In the previous discussion on the optimal control of merging maneuvers, the decision on forming a platoon had already been made. Next we discuss how such a decision-making process can take place and how an opportunistic platoon formation fits into the cooperation layer according to Figure~\ref{fig_architecture_upperlayers}. The aim of the opportunistic platoon formation is to decide whether it is fuel-efficient to form a platoon with a nearby heavy-duty vehicle and, if so, determine where the merge should take place to maximize the fuel savings.

A pairwise platoon formation strategy is proposed. Let $s_1^0$, $s_2^0$ denote the initial positions of a pair of vehicles and $s^f$ their common destination. The decision on whether to form a platoon will be based on the computation of the optimal merging point $s^m$. Contrary to the detailed merging maneuver in the previous section, the current platoon formation scenario is  performed over a potentially large geographical region and large distances. As a result, vehicle dynamics can be neglected and constant vehicle velocities $\bar{v}_i$ and platoon velocity $\bar{v}^p$ are assumed. This assumption additionally implies that no detailed road topography information is needed for this decision-making. Recall that the cooperative look-ahead controller is supposed to guarantee such average velocities even over roads with varying topography, whereas the merging controller will execute the actual merging maneuvers when the vehicles are sufficiently close.

The optimal fuel cost of forming a platoon will be compared to the fuel consumption of the two vehicles driving to their destination independently. As a result, only the effect of aerodynamic drag has to be considered and the average fuel flows follow from (\ref{eqn_forces_airdrag}) and (\ref{eqn_fuelflow}) as
\begin{align}
\bar{\varphi}_i &= \tfrac{1}{2} p_1 c_{\text{\rm d}}^0\rho A\bar{v}_i^3 + p_0, \quad i\in\{1,2\},\label{eqn_opf_fuelflow_vehicle}\\
\bar{\varphi}^p &= \tfrac{1}{2}p_1 \big( c_{\text{\rm d}}^0 + c_{\text{\rm d}}(\dtref) \big)\rho A (\bar{v}^p)^3 + 2p_0. \label{eqn_opf_fuelflow_platoon}
\end{align}
Here, $\bar{\varphi}_i$ gives the fuel flow of a vehicle without a predecessor as captured through the nominal air drag coefficient $c_{\text{\rm d}}^0$, while $\bar{\varphi}^p$ is the fuel flow of the two-vehicle platoon. Obviously, $\bar{\varphi}^p<\bar{\varphi}_1 + \bar{\varphi}_2$. The corresponding fuel cost now reads
\begin{align}
\bar{J}_{\textsc{opf}} = \sum_{i=1}^2 \bar{\varphi}_i\frac{s^m - s_i^0}{\bar{v}_i} + \bar{\varphi}^p\frac{s^f - s^m}{\bar{v}^p},
\label{eqn_opf_cost}
\end{align}
in which the merging point $s^m$ can be expressed as
\begin{align}
s^m = \frac{\bar{v}_2s_1^0 - \bar{v}_1s_2^0}{\bar{v}_2 - \bar{v}_1}.
\label{eqn_opf_mergingpoint}
\end{align}
Then, the fuel-optimal platoon formation problem can be stated as
\begin{equation}
\begin{aligned}
\min \quad & \text{Total average fuel consumption (\ref{eqn_opf_cost}),} \\
\text{subj. to} \quad
& \text{Constraint on the merging point (\ref{eqn_opf_mergingpoint}),} \\
& \text{Constraints on the average velocity,} \\
\end{aligned}
\nonumber
\end{equation}
in which the constraints on the average velocity are such that the platoon formation does not lead to a delayed arrival at the final destination $s^f$ nor that road speed limits are violated. If $\bar{J}_{\textsc{opf}}^*$ denotes the optimal solution, then a platoon is formed between the considered two vehicles if this total fuel cost is less than that of the two vehicles driving independently, i.e.,
\begin{align}
\bar{J}_{\textsc{opf}}^* < \sum_{i=1}^2 \bar{\varphi}_i \frac{s^f - s_i^0}{\bar{v}_{\text{nom},i}},
\end{align}
with $\bar{v}_{\text{nom},i}$ the nominal average velocity of vehicle $i$. Details on this opportunistic platoon formation can be found in~\cite{Liang2015}, whereas an alternative heuristic approach is given in~\cite{larson_2015}.

\section{Fleet management layer}\label{sec_fleetmanagementlayer}

\subsection{Fleet management architecture}\label{sec_fleetmanagementarchitecture}

\begin{figure}
\begin{center}
\begin{tikzpicture}[
  separator/.style={thin,dashed},
  box/.style={rectangle, draw, align=center, inner sep=0mm, minimum height=12mm, minimum width=70mm},
  signal/.style={semithick,>=stealth},
]
\node (planning) at (0mm,20mm) [box] {Transport planning};
\node (routing) at (0mm,0mm) [box] {Routing};
\node (coordination) at (0mm,-20mm) [box] {Coordinated platoon planning};
\draw[<-,signal] ([xshift=-2mm]planning.south) to ([xshift=-2mm]routing.north);
\draw[->,signal] ([xshift= 2mm]planning.south) to ([xshift= 2mm]routing.north);
\draw[<-,signal] ([xshift=-2mm]routing.south) to ([xshift=-2mm]coordination.north);
\draw[->,signal] ([xshift= 2mm]routing.south) to ([xshift= 2mm]coordination.north);
\node at (-41mm,0mm) [align=center, rotate=90] {Fleet layer};
\draw[-,separator] (-37mm-6mm,-30mm) to (37mm,-30mm);
\draw[-,separator] (-37mm-6mm,30mm) to (37mm,30mm);
\node at (0mm,-33mm) [align=center] {Cooperation layer};
\draw[<-,signal,dashed] ([xshift=-2mm]coordination.south) to (-2mm,-30mm);
\draw[->,signal,dashed] ([xshift= 2mm]coordination.south) to (2mm,-30mm);
\end{tikzpicture}
\caption{Fleet management architecture according to the fleet layer in Figure~\ref{fig_architecture}.}
\label{fig_architecture_fleetmanagement}
\end{center}
\end{figure}
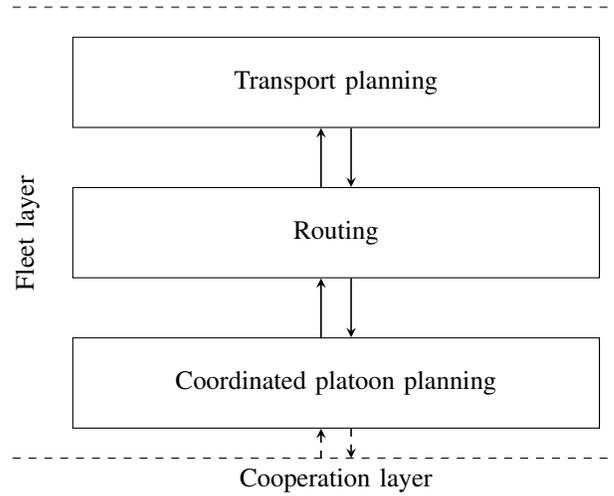

The fleet management layer handles transport planning, routing and coordination, as detailed in Figure~\ref{fig_architecture_fleetmanagement}. Transport planning amounts to distributing the flow of goods over the available vehicles in the fleet. This is a logistics problem in which the available resources are managed to meet customer requirements. The assignment of goods to vehicles is optimized by combining similar assignments to the same vehicle. Size, weight, and type of cargo must be considered. The availability of drivers and the drivers' legal resting times are other parameters that should be regarded.

Routing is the process of finding the most suitable path from the origin to the destination. In our setting the aim is to find the most fuel-efficient route. The topography of the road has a large influence on the fuel consumption, in particular for heavy-duty vehicles. The traffic conditions, estimated from historic and real-time data, and current and predicted weather should also be taken into account. Equally important is the reliability of the plan, as accurate predictions of the time of arrival and of the corresponding fuel consumption are essential.

The platoon coordination jointly adjust the motion along the vehicles' paths. Of particular interest is the ability to adjust the velocity profiles in order to form fuel-efficient platoons. A procedure for such coordinated platoon planning is described in the following section.

\subsection{Coordinated platoon planning}\label{sec_coordinatedplatoonplanning}
The modern communication infrastructure allows for the fusion of real-time position, velocity, and assignment information of heavy-duty vehicles together with external influences such as traffic data and thus enables a centralized coordination of a large number of vehicles over great distances. In this section, a method for the coordination of a potentially large fleet of vehicles is described, aimed at achieving fuel savings through the formation of platoons.
This approach can be regarded as an extension of the opportunistic platoon formation approach of Section~\ref{sec_opportunisticplatoonformation}, where the latter is inherently local in nature.

In order to efficiently obtain platoon configurations and the corresponding average velocities for each vehicle, a three-step approach is taken. The first step comprises finding the most suitable route for each vehicle, taking factors such as road topography and traffic information into account.

Second, for a given vehicle, which will be referred to as a coordination leader, its fuel-optimal velocity profile is computed. The starting time and arrival deadline are taken into account together with constraints such as driver resting times. The profile specifies the desired average velocity, which will later be refined by the use of cooperative look-ahead control. Then, for each vehicle with a partially overlapping route with the coordination leader, the pairwise analysis of Section~\ref{sec_opportunisticplatoonformation} is used to determine whether it is beneficial to adapt its velocity profile to form or join a platoon with the coordination leader. If so, this vehicle is referred to as a coordination follower. In this pairwise analysis, the coordination leader does not adapt its velocity profile, such that several coordination followers can be assigned to a single coordination leader. Arrival deadlines are taken into account when adapting the velocity profiles of the coordination followers.

The selection of the most suitable coordination leaders is crucial in obtaining significant fuel savings. This selection forms the third step. Repeating the pairwise analysis for every potential coordination leader leads to a data set that can be conveniently represented as a graph. In this graph, the nodes represent the vehicles and their incoming edges denote the fuel savings obtained when this vehicle is selected as a coordination leader.
From graph clustering algorithms~\cite{pam_book}), an algorithm can be derived to compute a suitable set of coordination leaders. Specifically, a greedy algorithm that incrementally adds or removes individual vehicles from the set of coordination leaders provides a computationally efficient and scalable
approach~\cite{vandehoef_2015}. Instead of coordination of vehicles through adaptation of their velocity profiles, vehicle sorting for platooning has been considered~\cite{Hall_platoon_sorting}, as well as other platooning algorithms~\cite{datamining_platooning}.

\subsection{Incentives for cooperation}
There are many incentives for individual owners of truck fleets to optimize their long-haulage transportation tasks. By coordinating timing and routing of vehicles, the fleet owner can utilize their available resources (fuel, vehicles, drivers, etc.) as efficiently as possible. Through vehicle platooning, the tasks can be further optimized and fuel consumption decreased, as discussed in this paper. The long-haulage transport and logistics industry consists of a large and diverse set of fleet owners, however, and it is for obvious reasons hard for many of them to cooperate without financial guarantees and trust. To be able to capitalize on vehicle cooperation, we need to have as big pool as possible of heavy-duty vehicles that travel on the same (or similar) route and at the same time. It is rarely the case for small fleet owners to have so many similar tasks to satisfy this criterion. One solution to this problem is instead to create a fleet management service for the owners and their vehicles. In such a service, the fleet owners can privately provide their routes and timetables so that the service provider can pair the vehicles for cooperation. For participating in this service, the fleet owners may need to pay a subscription fee in addition to invest in devices to facilitate cooperation. The case of cooperative heavy-vehicle platooning is discussed next.

A fleet management service for heavy-duty vehicle cooperation focusing on platooning can be evaluated considering existing patterns of long-haulage goods delivery. Based on position data from thousands of heavy-duty vehicles, it has been shown that many vehicles have other vehicles in their vicinity, even when only a single vehicle brand is considered~\cite{liang_2014}. Hence, by simply slowing down a bit or speeding up, it is possible with minimum effort to form a vehicle platoon, as was described in Section~\ref{sec_cooperationlayer}. It is also clear from these data that quite a few vehicles are actually driving in spontaneous platoons already today. To automate a platooning service, it is essential to present transparent information on benefits and costs to individual fleet owners and drivers. By utilizing economic theory on technology adoption~\cite{rei83} and data from actual transportation tasks~\cite{liang_2014}, it is possible to reason how a market for such a service can be established~\cite{far+15msc}. One example is centralized cooperation, in which fleet owners pay to subscribe to a third-party service provider and then can cooperate with any other fleet owner who is part of the system. The pricing strategy needs to be carefully developed for such a service, as the marginal benefit for joining such a system for a large fleet owner might be smaller than for a fleet owner with few vehicles.

\section{Case study}\label{sec_evaluation}

\subsection{Scenario}

\begin{figure*}
\begin{center}
\def\svgwidth{1.35\columnwidth}
\footnotesize
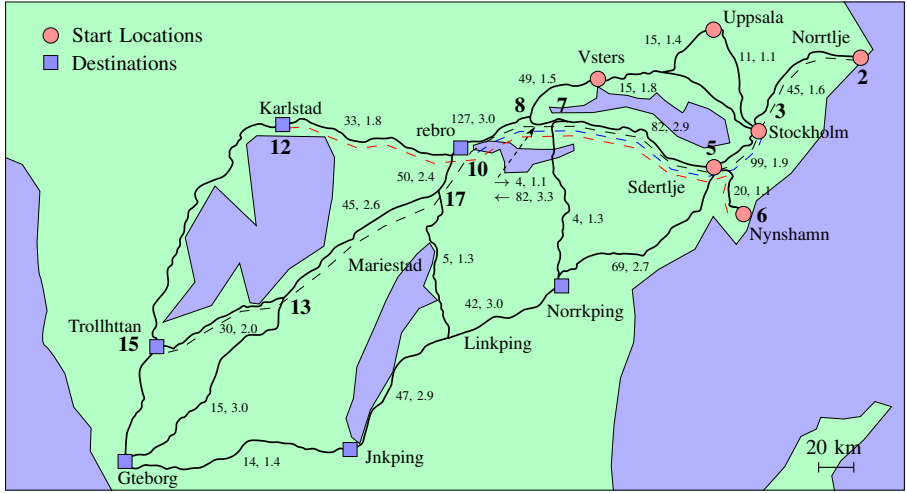
\caption{The Swedish road network used in the case study. Starting locations and destinations are indicated by red circles and blue squares, respectively, and the boldface numbers represent intersections. The two numbers next to the road segments indicate the number of vehicles that traverse this segment and the average platoon size on this segment, respectively, as a result of the applied coordination algorithm. The road segment between nodes 7 and 8 is traversed in both directions and the statistics for vehicles travelling in either direction are indicated separately. Three routes, indicated by dashed lines, are   highlighted as an example for a group comprising a coordination leader (black) and two coordination followers (blue and red).}
\label{fig_abstracted_network}
\end{center}
\end{figure*}

The platoon control and coordination algorithms presented in this paper are demonstrated by means of a simulation scenario representing a part of the highway network of Sweden, see Figure~\ref{fig_abstracted_network}. On this network, 200 heavy-duty vehicles originating from six locations in the Stockholm area in the east travel to five destinations in the west. The starting times for these vehicles are taken from a two-hour interval, whereas the parameter values for each vehicle are given in Table~\ref{tab_modelparameters}. The coordinated platoon planning for this scenario is evaluated in Section~\ref{sec_evaluation_platoonplanning} before focusing on the cooperative look-ahead control for one specific platoon in Section~\ref{sec_evaluation_clac}.

\begin{table}
\begin{center}
  \caption{\rm Model parameters used in the experimental evaluation.}
  \label{tab_modelparameters}
  \begin{tabular}{|ccc||ccc|}
  \hline
  $m$ & $40\,000$ & kg               &  $p_0$ & $5.36\cdot10^{-4}$ & kg\,s$^{-1}$ \\
  $A$ & $10$ & m$^2$               &  $p_1$ & $5.15\cdot10^{-8}$ & kg\,s$^{-1}$\,W$^{-1}$ \\
  $c_{\text{\rm d}}^0$ & $0.6$ & &  $\rho$ & $1.29$ & kg\,m$^{-3}$ \\
  $\alpha_1$ & $0.53$ &           &  $P_\text{min}$ & $-9$ & kW \\
  $\alpha_2$ & $0.81$ & s$^{-1}$   &  $P_\text{max}$ & $300$ & kW \\
  \hline
  \end{tabular}
\end{center}
\end{table}

\subsection{Coordinated platoon planning evaluation}\label{sec_evaluation_platoonplanning}
The methodology for coordinated platoon planning in Section~\ref{sec_coordinatedplatoonplanning} is used to select suitable coordination leaders and their respective followers. Herein, pairwise plans are considered in which the coordination followers catch up with their leaders and platoon until either of their routes end or their routes split up.

For this scenario, the coordination algorithm selects 54 coordination leaders and 139 coordination followers, which adjust their velocity profiles to catch up with the coordination leaders in order to form platoons. The maximum number of coordination followers per coordination leader is 8, whereas the median is 2. The remaining 7 vehicles do not platoon but traverse their routes individually.

The coordinated platoon planning amounts to a fuel saving of $5.7$\,\%, when compared to all vehicles driving independently. Considering that the maximum fuel saving is $12.0$\,\% when vehicles platoon continuously, the coordination layer is fairly efficient in this scenario. Recall that the velocity adjustments necessary for coordination lead to an increased fuel consumption. The total fuel savings amount to $1\,045$ liters of diesel fuel and a reduction of CO$_2$ emissions of $2\,770$\,kg.

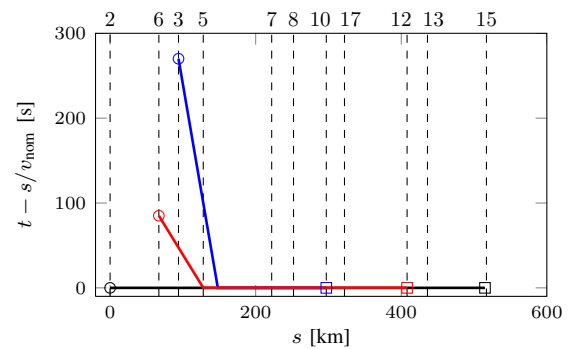
\begin{figure}
\begin{center}
\begin{tikzpicture}
\begin{axis}[clip=false,enlargelimits=false,
  height=35mm,
  width=60mm,
  at={(0,0)}, scale only axis,
  xlabel={$s$ [km]},
  ylabel={$t - s/v_{\text{nom}}$ [s]},
  xmin=-20, xmax=600, ymin=-10, ymax=300,
  x label style={font=\footnotesize, yshift=1mm},
  y label style={font=\footnotesize},
  tick label style={font=\scriptsize},
]
\foreach \x/\l in {0/2,67/6,94/3,128/5,222/7,252/8,517/15}{
   \edef\temp{\noexpand%
     \draw[dashed,black] (axis cs:\x,-10) -- (axis cs:\x,300) node[pos=1,above,scale=0.8] {$\l$};
   }
   \temp
}
\draw[dashed,black] (axis cs:297,-10) -- (axis cs:297,300) node[pos=1,above,scale=0.8,xshift=-1mm] {$10$};
\draw[dashed,black] (axis cs:322,-10) -- (axis cs:322,300) node[pos=1,above,scale=0.8,xshift=1mm] {$17$};
\draw[dashed,black] (axis cs:408,-10) -- (axis cs:408,300) node[pos=1,above,scale=0.8,xshift=-1mm] {$12$};
\draw[dashed,black] (axis cs:436,-10) -- (axis cs:436,300) node[pos=1,above,scale=0.8,xshift=1mm] {$13$};
\addplot[black,line width=1pt] coordinates { (0,0) (515,0) };
\addplot[black,mark=o] coordinates { (0,0) };
\addplot[black,mark=square] coordinates { (515,0) };
\addplot[blue,line width=1pt] coordinates { (94,270) (148,0) (297,0) };
\addplot[blue,mark=o] coordinates { (94,270) };
\addplot[blue,mark=square] coordinates { (297,0) };
\addplot[red,line width=1pt] coordinates { (67,85) (128,0) (297,0) (408,0) };
\addplot[red,mark=o] coordinates { (67,85) };
\addplot[red,mark=square] coordinates { (408,0) };
\end{axis}
\end{tikzpicture}
\vskip-2mm%
\caption{Platoon plans for a coordination leader (black) and two coordination followers (blue and red) that catch up with the coordination leader to form a platoon. The graph shows the time gap to the platoon leader as a function of the position on the road, where this position is taken along the routes of the individual vehicles. The dashed lines denote the position of the nodes representing road intersections in Figure~\ref{fig_abstracted_network}, with the top labels denoting the node number. As an example, note that the coordination leader starts from Norrt\"{a}lje (node 2) and drives to Trollh\"{a}ttan (node 15). When the time gap is zero and the routes of the vehicles overlap (between nodes 5 and 10), the vehicles operate in a platoon.}
\label{fig:adaptation_trajectory.pdf}
\end{center}
\end{figure}

The routes of one particular coordination leader and its two coordination followers are highlighted in Figure~\ref{fig_abstracted_network}. The corresponding trajectories are presented in Figure~\ref{fig:adaptation_trajectory.pdf}, where the time gaps with respect to the coordination leader as a function of the position on the road are shown. Note that the first coordination follower (blue) shares the first part of its route with the coordination leader (black), but as it starts 1.25 hours later it catches up at maximum speed, indicated by a decreasing gap to the leader in Figure~\ref{fig:adaptation_trajectory.pdf}. It then meets the platoon consisting of the coordination leader and the other coordination follower (red) between nodes 5 and 7, in which it stays until its destination at node 10 is reached. The route of the second coordination follower intersects with the route of the coordination leader at node 5 and the coordination follower's start time is such that it catches up to the coordination leader at a velocity that is lower than the maximum speed. The coordination follower and coordination leader form a platoon at node 5 and platoon until node 10 where their routes split~up.

\subsection{Cooperative look-ahead control evaluation} \label{sec_evaluation_clac}

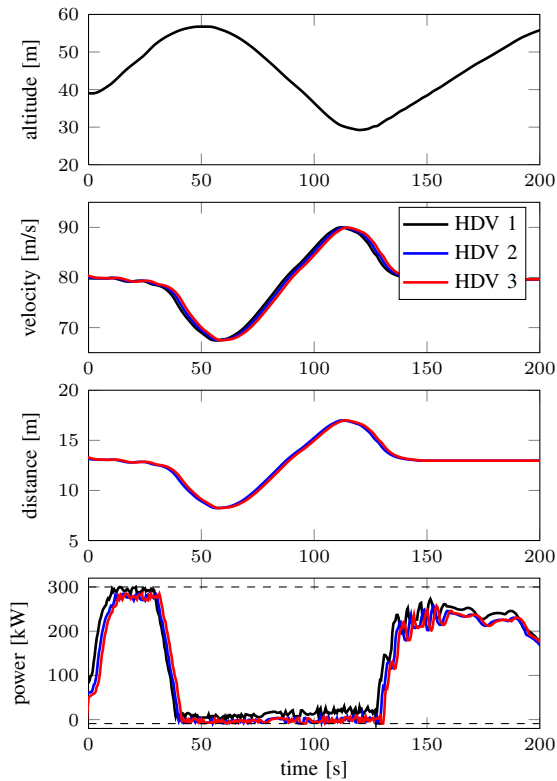
\begin{figure}
\begin{center}
\begin{tikzpicture}
\begin{axis}[
  height=20mm,
  width=60mm,
  at={(0,0)}, scale only axis,
  xlabel={},
  ylabel={altitude [m]},
  xmin=0, xmax=200, ymin=20, ymax=60,
  x label style={font=\footnotesize, yshift=1mm},
  y label style={font=\footnotesize},
  tick label style={font=\scriptsize},
]
\addplot[black,line width=1pt] table[x index=0,y index=3] {img/tikzdata_clac_local_hdv1.dat};
\end{axis}
\begin{axis}[
  height=20mm,
  width=60mm,
  at={(0,-25mm)}, scale only axis,
  xlabel={},
  ylabel={velocity [m/s]},
  xmin=0, xmax=200, ymin=65, ymax=95,
  x label style={font=\footnotesize, yshift=1mm},
  y label style={font=\footnotesize},
  tick label style={font=\scriptsize},
  legend style={legend pos=north east,font=\footnotesize, inner sep=1pt},
]
\addplot[black,line width=1pt] table[x index=0,y index=1] {img/tikzdata_clac_local_hdv1.dat};
\addlegendentry{HDV 1};
\addplot[blue,line width=1pt] table[x index=0,y index=1] {img/tikzdata_clac_local_hdv2.dat};
\addlegendentry{HDV 2};
\addplot[red,line width=1pt] table[x index=0,y index=1] {img/tikzdata_clac_local_hdv3.dat};
\addlegendentry{HDV 3};
\end{axis}
\begin{axis}[
  height=20mm,
  width=60mm,
  at={(0,-50mm)}, scale only axis,
  xlabel={},
  ylabel={distance [m]},
  xmin=0, xmax=200, ymin=5, ymax=20,
  x label style={font=\footnotesize, yshift=1mm},
  y label style={font=\footnotesize},
  tick label style={font=\scriptsize},
]
\addplot[blue,line width=1pt] table[x index=0,y index=3] {img/tikzdata_clac_local_hdv2.dat};
\addplot[red,line width=1pt] table[x index=0,y index=3] {img/tikzdata_clac_local_hdv3.dat};
\end{axis}
\begin{axis}[
  height=20mm,
  width=60mm,
  at={(0,-75mm)}, scale only axis,
  xlabel={time [s]},
  ylabel={power [kW]},
  xmin=0, xmax=200, ymin=-20, ymax=320,
  x label style={font=\footnotesize, yshift=1mm},
  y label style={font=\footnotesize},
  tick label style={font=\scriptsize},
]
\addplot[black,line width=1pt] table[x index=0,y index=2] {img/tikzdata_clac_local_hdv1.dat};
\addplot[blue,line width=1pt] table[x index=0,y index=2] {img/tikzdata_clac_local_hdv2.dat};
\addplot[red,line width=1pt] table[x index=0,y index=2] {img/tikzdata_clac_local_hdv3.dat};
\addplot[black,dashed] coordinates {(0,-8.9945) (200,-8.9945)};
\addplot[black,dashed] coordinates {(0,300) (200,300)};
\end{axis}
\end{tikzpicture}
\vskip-2mm%
\caption{Cooperative look-ahead control Local behavior of the three-vehicle platoon depicted in Figure~\ref{fig:adaptation_trajectory.pdf}. The plots show the road topography experienced by the leading vehicle, the vehicle speeds, the inter-vehicle distances, and the generated power, respectively. The generated power is the sum of the engine power and the power dissipated by the braking system. The dashed lines represent the minimum and maximum engine powers.}
\label{fig:platon_layer_simulation}
\end{center}
\end{figure}

The cooperative look-ahead control and the vehicle layer govern the local behavior of each platoon by explicitly taking into account topography information and traffic. Figure~\ref{fig:platon_layer_simulation} illustrates the effective behavior of the three-vehicle platoon discussed in the previous section when driving along a $4$\,km road stretch in the latter part of the segment between node 5 and node 7. It can be observed that the three vehicles follow approximately the same velocity profile, albeit shifted in time as required by their cooperative look-ahead control strategy. This translates into the vehicles following approximately the same velocity profile in the spatial domain and, due to the dependence of the slope on position, results in similar power profiles.

In order to respond to the fuel-optimality criterion, the cooperative look-ahead control requires the vehicles to follow a particular speed profile depending on the road topography. Specifically, it requires the vehicles to keep a constant speed of $80$\,km/h during the uphill segment and to drop the speed down to $68$\,km/h at the top of the hill. This allows the vehicles to gain speed during the downhill without reaching the speed limit of $90$\,km/h. In particular, the required downhill speed profile is such that the lead vehicle fuels slightly, whereas the follower vehicles coast (i.e., they do not fuel). Hereby, the desired inter-vehicular distances are maintained even though the follower vehicles experience a reduced aerodynamic drag. Hence, the proposed cooperative controller avoids braking and exploits the combined potential of both platooning and look-ahead control.

The cooperative look-ahead control combines the potential of platooning and look-ahead control and achieves larger fuel savings than for each of the methods independently. For the hilly stretch shown in Figure \ref{fig:platon_layer_simulation}, it allows to save approximately $10$\% of energy compared to the vehicles driving alone using look-ahead control and $7$\% compared to the vehicles platooning without cooperating and exploiting topography information.

\section{Conclusions}\label{sec_conclusions}
A cyber-physical systems approach towards the control and coordination of a large-scale transportation system was presented in this paper. The approach relies on modern vehicle-to-vehicle and vehicle-to-infrastructure communication and is supported by ubiquitous computation power as offered through cloud services and onboard computers. The coordination of heavy-duty vehicles is aimed at the reduction of fuel consumption and a layered freight transport system architecture was developed that achieved this reduction through exploiting the formation of closely-spaced groups of vehicles, which experienced a reduced aerodynamic drag. The distributed control of platooning vehicles was handled in the low-level vehicle layer of the system architecture, whereas the middle-level cooperation layer
employed look-ahead control to further reduce fuel consumption. The formation of platoons was also handled in this cooperation layer. Finally, the fleet layer on top performed the large-scale coordination of the platoons with integrated routing and transport planning. This allowed both small and large fleet owners to benefit from the fuel-saving potential of cooperation. A case study involving 200 vehicles confirmed the feasibility of this cyber-physical approach to freight transport.

Extensive real-world experimental evaluation of the approach developed in the paper is the scope of future work. Such evaluation should include both small- and large-scale tests. Experiments with vehicles on public roads are obviously needed to study many practical implications. Such experiments can build on earlier experiences of individual platoon experiments on Swedish roads~\cite{alam_2010,alam_2015b}.

\section*{Acknowledgements}
The work presented in this paper has greatly benefited from a long-term collaboration between KTH and Scania. We would in particular like to acknowledge important contributions by Magnus Adolfsson, Henrik Pettersson, and Tony Sandberg. Funding is received from the European Union Seventh Framework Programme under the project COMPANION, Sweden's innovation agency VINNOVA-FFI, the Knut and Alice Wallenberg Foundation, and the Swedish Research Council.


\bibliographystyle{plain}
\bibliography{bibliography_total}

\begin{thebibliography}{10}

\bibitem{gharavi_2007}
Special issue on advanced automobile technologies.
\newblock {\em Proceedings of the IEEE}, 95(2), 2007.

\bibitem{buehler_2008}
Special issue on the {DARPA} {U}rban {C}hallenge.
\newblock {\em Journal of Field Robotics}, 25(8-9), 2008.

\bibitem{alam_2015b}
A.~Alam, B.~Besselink, V.~Turri, J.~M{\aa}rtensson, and K.H. Johansson.
\newblock Heavy-duty vehicle platooning towards sustainable freight
  transportation.
\newblock 2015 (submitted).

\bibitem{alam_2010}
A.~Alam, A.~Gattami, and K.H. Johansson.
\newblock An experimental study on the fuel reduction potential of heavy duty
  vehicle platooning.
\newblock In {\em Proceedings of the 13th International IEEE Conference on
  Intelligent Transportation Systems, Madeira, Portugal}, pages 306--311, 2010.

\bibitem{alam_2013}
A.~Alam, J.~M{\aa}rtensson, and K.H. Johansson.
\newblock Look-ahead cruise control for heavy duty vehicle platooning.
\newblock In {\em Proceedings of the 16th International IEEE Annual Conference
  on Intelligent Transportation Systems, The Hague, The Netherlands}, pages
  928--935, 2013.

\bibitem{armbrust_2010}
M.~Armbrust, A.~Fox, R.~Griffith, A.D. Joseph, R.~Katz, A.~Konwinski, G.~Lee,
  D.~Patterson, A.~Rabkin, I.~Stoica, and M.~Zaharia.
\newblock A view of cloud computing.
\newblock {\em Communications of the ACM}, 53(4):50--58, 2010.

\bibitem{beard_2001}
R.W. Beard, J.~Lawton, and F.Y. Hadaegh.
\newblock A coordination architecture for spacecraft formation control.
\newblock {\em IEEE Transactions on Control Systems Technology}, 9(6):777--790,
  2001.

\bibitem{book_bellman_1957}
R.~Bellman.
\newblock {\em Dynamic programming}.
\newblock Princeton University Press, New Jersey, USA, 1957.

\bibitem{bengler_2014}
K.~Bengler, K.~Dietmayer, B.~Farber, M.~Maurer, C.~Stiller, and H.~Winner.
\newblock Three decades of driver assistance systems: review and future
  perspectives.
\newblock {\em IEEE Intelligent Transportation Systems Magazine}, 6(4):6--22,
  2014.

\bibitem{besselink_2015b}
B.~Besselink and K.H. Johansson.
\newblock String stability and a delay-based spacing policy for vehicle
  platoons subject to disturbances.
\newblock 2015 (submitted).

\bibitem{bonnet_2000}
C.~Bonnet and H.~Fritz.
\newblock Fuel consumption reduction in a platoon: {E}xperimental results with
  two electronically coupled trucks at close spacing.
\newblock In {\em Proceedings of the Future Transportation Technology
  Conference, Costa Mesa, USA}, SAE Technical Paper 2000-01-3056, 2000.

\bibitem{chandler_1958}
R.E. Chandler, R.~Herman, and E.W. Montroll.
\newblock Traffic dynamics: {S}tudies in car following.
\newblock {\em Operations Research}, 6(2):165--184, 1958.

\bibitem{chu_1974}
K.-C. Chu.
\newblock Decentralized control of high-speed vehicular strings.
\newblock {\em Transportation Science}, 8(4):361--384, 1974.

\bibitem{dickmanns_1994}
E.D. Dickmanns, R.~Behringer, D.~Dickmanns, T.~Hildebrandt, M.~Maurer,
  F.~Thomanek, and J.~Schiehlen.
\newblock The seeing passenger car {VaMoRs-P}.
\newblock In {\em Proceedings of the IEEE Intelligent Vehicles Symposium,
  Paris, France}, pages 68--73, 1994.

\bibitem{eu_pocketbook_2014}
{European Commission}.
\newblock \textit{{EU} transport in figures -- {S}tatistical pocketbook}.
\newblock Publications Office of the European Union, Luxembourg, 2014.

\bibitem{far+15msc}
F.~Farokhi, K.-Y. Liang, and K.H. Johansson.
\newblock Cooperation patterns between fleet owners for transport assignments.
\newblock In {\em Proceedings of the IEEE Multi-Conference on Systems and
  Control, Sydney, Australia}, 2015 (submitted).

\bibitem{fenton_1968}
R.E. Fenton, R.L. Cosgriff, K.~Olson, and L.M. Blackwell.
\newblock One approach to highway automation.
\newblock {\em Proceedings of the IEEE}, 56(4):556--566, 1968.

\bibitem{gm_film_1940}
{General Motors}.
\newblock \textit{To new horizons}.
\newblock Film, available online at http://youtu.be/aIu6DTbYnog?t=14m27s, 1940.

\bibitem{Hall_platoon_sorting}
R.~Hall and C.~Chin.
\newblock Vehicle sorting for platoon formation: {I}mpacts on highway entry and
  throughput.
\newblock {\em Transportation Research Part C: Emerging Technologies},
  13(5-6):405--420, 2005.

\bibitem{hartenstein_2008}
H.~Hartenstein and K.P. Laberteaux.
\newblock A tutorial survey on vehicular ad hoc networks.
\newblock {\em IEEE Communications Magazine}, 46(6):164--171, 2008.

\bibitem{hellstrom_2009}
E.~Hellstr\"{o}m, M.~Ivarsson, J.~{\AA}slund, and L.~Nielsen.
\newblock Look-ahead control for heavy trucks to minimize trip time and fuel
  consumption.
\newblock {\em Control Engineering Practice}, 17(2):245--254, 2009.

\bibitem{herrera+10}
J.C. Herrera, D.B. Work, R.~Herring, X.~Ban, Q.~Jacobson, and A.M. Bayen.
\newblock Evaluation of traffic data obtained via {GPS}-enabled mobile phones:
  {T}he {M}obile {C}entury field experiment.
\newblock {\em Transportation Research Part C: Emerging Technologies},
  18(4):568--583, 2010.

\bibitem{hooker1988optimal}
J.N. Hooker.
\newblock Optimal driving for single-vehicle fuel economy.
\newblock {\em Transportation Research Part A: General}, 22(3):183--201, 1988.

\bibitem{horowitz_2000}
R.~Horowitz and P.~Varaiya.
\newblock Control design of an automated highway system.
\newblock {\em Proceedings of the IEEE}, 88(7):913--925, 2000.

\bibitem{book_hucho_1998}
W.-H. Hucho, editor.
\newblock {\em Aerodynamics of road vehicles}.
\newblock Society of Automotive Engineers, USA, 4th edition, 1998.

\bibitem{ioannou_1993}
P.A. Ioannou and C.C. Chien.
\newblock Autonomous intelligent cruise control.
\newblock {\em IEEE Transactions on Vehicular Technology}, 42(4):657--672,
  1993.

\bibitem{johansson_2005}
K.H. Johansson, M.~T\"{o}rngren, and L.~Nielsen.
\newblock Vehicle applications of controller area network.
\newblock In D.~Hristu-Varsakelis and L.S. Levine, editors, {\em Handbook of
  Networked and Embedded Control Systems}, pages 741--765. Birkh\"{a}user,
  Boston, USA, 2005.

\bibitem{karagiannis_2011}
G.~Karagiannis, O.~Altintas, E.~Ekici, G.~Heijenk, B.~Jarupan, K.~Lin, and
  T.~Weil.
\newblock Vehicular networking: {A} survey and tutorial on requirements,
  architectures, challenges, standards and solutions.
\newblock {\em IEEE Communications Surveys~\& Tutorials}, 13(4):584--616, 2011.

\bibitem{pam_book}
L.~Kaufman and P.J. Rousseeuw.
\newblock {\em Finding groups in data: {A}n introduction to cluster analysis}.
\newblock John Wiley \& Sons, Hoboken, USA, 2008.

\bibitem{koller_2015}
J.P.J. Koller, A.~Grossmann~Col\'{\i}n, B.~Besselink, and K.H. Johansson.
\newblock Fuel-efficient control of merging maneuvers for heavy-duty vehicle
  platooning.
\newblock In {\em Proceedings of the 18th IEEE International Conference on
  Intelligent Transportation Systems, Las Palmas, Spain}, 2015 (submitted).

\bibitem{lammert_2014}
M.P. Lammert, A.~Duran, J.~Diez, K.~Burton, and A.~Nicholson.
\newblock Effect of platooning on fuel consumption of class 8 vehicles over a
  range of speeds, following distances, and mass.
\newblock {\em SAE International Journal of Commercial Vehicles},
  7(2):626--639, 2014.

\bibitem{larson_2015}
J.~Larson, K.-Y. Liang, and K.H. Johansson.
\newblock A distributed framework for coordinated heavy-duty vehicle
  platooning.
\newblock {\em IEEE Transactions on Intelligent Transportation Systems},
  16(1):419--429, 2015.

\bibitem{levine_1966}
W.~Levine and M.~Athans.
\newblock On the optimal error regulation of a string of moving vehicles.
\newblock {\em IEEE Transactions on Automatic Control}, AC-11(3):355--361,
  1966.

\bibitem{liang_2014}
K.-Y. Liang, J.~M{\aa}rtensson, and K.H. Johansson.
\newblock Fuel-saving potentials of platooning evaluated through sparse
  heavy-duty vehicle position data.
\newblock In {\em Proceedings of the IEEE Intelligent Vehicles Symposium,
  Dearborn, USA}, pages 1061--1068, 2014.

\bibitem{Liang2015}
K.-Y. Liang, J.~M{\aa}rtensson, and K.H. Johansson.
\newblock Fuel-efficient forming of heavy-duty vehicle platoons.
\newblock {\em IEEE Transactions on Intelligent Transportation Systems}, 2015
  (submitted).

\bibitem{datamining_platooning}
P.~Meisen, T.~Seidl, and K.~Henning.
\newblock A data-mining technique for the planning and organization of truck
  platoons.
\newblock In {\em Proceedings of the International Conference on Heavy
  Vehicles, Paris, France}, pages 389--402, 2008.

\bibitem{melzer_1971}
S.M. Melzer and B.~Kuo.
\newblock A closed-form solution for the optimal error regulation of a string
  of moving vehicles.
\newblock {\em IEEE Transactions on Automatic Control}, 16(1):50--52, 1971.

\bibitem{monastyrsky1993rapid}
V.V. Monastyrsky and I.M. Golownykh.
\newblock Rapid computation of optimal control for vehicles.
\newblock {\em Transportation Research Part B: Methodological}, 27(3):219--227,
  1993.

\bibitem{naus_2010}
G.J.L. Naus, R.P.A. Vugts, J.~Ploeg, M.J.G. Van~de Molengraft, and
  M.~Steinbuch.
\newblock String-stable {CACC} design and experimental validation: a
  frequency-domain approach.
\newblock {\em IEEE Transactions on Vehicular Technology}, 59(9):4268--4279,
  2010.

\bibitem{itf_transportoutlook_2015}
{OECD/International Transport Forum}.
\newblock \textit{International transport outlook 2015}.
\newblock OECD Publishing/ITF, 2015.

\bibitem{peppard_1974}
L.E. Peppard.
\newblock String stability of relative-motion {PID} vehicle control systems.
\newblock {\em IEEE Transactions on Automatic Control}, 19(5):579--581, 1974.

\bibitem{ploeg_2014}
J.~Ploeg, N.~van~de Wouw, and H.~Nijmeijer.
\newblock {$\mathcal{L}_p$} string stability of cascaded systems: application
  to vehicle platooning.
\newblock {\em IEEE Transactions on Control Systems Technology},
  22(2):786--793, 2014.

\bibitem{raza_1996}
H.~Raza and P.~Ioannou.
\newblock Vehicle following control design for automated highway systems.
\newblock {\em IEEE Control Systems Magazine}, 16(6):43--60, 1996.

\bibitem{rei83}
J.F. Reinganum.
\newblock Technology adoption under imperfect information.
\newblock {\em The Bell Journal of Economics}, 14(1):57--69, 1983.

\bibitem{scania_annualreport_2014}
{Scania AB}.
\newblock \textit{Annual report}, 2014.

\bibitem{schwarzkopf_1977}
A.B. Schwarzkopf and R.B. Leipnik.
\newblock Control of highway vehicles for minimum fuel consumption over varying
  terrain.
\newblock {\em Transportation Research}, 11(4):279--286, 1977.

\bibitem{seiler_2004}
P.~Seiler, A.~Pant, and K.~Hedrick.
\newblock Disturbance propagation in vehicle strings.
\newblock {\em IEEE Transactions on Automatic Control}, 49(10):1835--1842,
  2004.

\bibitem{shaikh_2007}
M.S. Shaikh and P.E. Caines.
\newblock On the hybrid optimal control problem: {T}heory and algorithms.
\newblock {\em IEEE Transactions on Automatic Control}, 52(9):1587--1603, 2007.

\bibitem{sheikholeslam_1993}
S.~Sheikholeslam and C.A. Desoer.
\newblock Longitudinal control of a platoon of vehicles with no communication
  of lead vehicle information: a system level study.
\newblock {\em IEEE Transactions on Vehicular Technology}, 42(4):546--554,
  1993.

\bibitem{stoicescu1995fuel}
A.P. Stoicescu.
\newblock On fuel-optimal velocity control of a motor vehicle.
\newblock {\em International Journal of Vehicle Design}, 16(2/3):229--256,
  1995.

\bibitem{sussmann_1999}
H.~Sussmann.
\newblock A maximum principle for hybrid optimal control problems.
\newblock In {\em Proceedings of the 38th IEEE Conference on Decision and
  Control, Phoenix, USA}, pages 425--430, 1999.

\bibitem{swaroop_1996}
D.~Swaroop and J.K. Hedrick.
\newblock String stability of interconnected systems.
\newblock {\em IEEE Transactions on Automatic Control}, 41(3):349--357, 1996.

\bibitem{swaroop_1994}
D.~Swaroop, J.K. Hedrick, C.C. Chien, and P.~Ioannou.
\newblock A comparison of spacing and headway control laws for automatically
  controlled vehicles.
\newblock {\em Vehicle System Dynamics}, 23(1):597--625, 1994.

\bibitem{tsugawa_2000}
S.~Tsugawa, S.~Kato, T.~Matsui, H.~Naganawa, and H.~Fujii.
\newblock An architecture for cooperative driving of automated vehicles.
\newblock In {\em Proceedings of the Intelligent Transportation Systems
  Conference, Dearborn, USA}, pages 422--427, 2000.

\bibitem{turri_2015}
V.~Turri, B.~Besselink, and K.H. Johansson.
\newblock Cooperative look-ahead control for fuel-efficient and safe heavy-duty
  vehicle platooning.
\newblock {\em IEEE Transactions on Control Systems Technology}, 2015
  (submitted).

\bibitem{vandehoef_2015}
S.H. van~de Hoef, K.H. Johansson, and D.V. Dimarogonas.
\newblock Fuel-optimal coordination of truck platooning based on shortest
  paths.
\newblock In {\em Proceedings of the American Control Conference, Chicago,
  USA}, 2015.

\bibitem{varaiya_1993}
P.~Varaiya.
\newblock Smart cars on smart roads: problems of control.
\newblock {\em IEEE Transactions on Automatic Control}, 38(2):195--207, 1993.

\bibitem{wang+12}
P.~Wang, T.~Hunter, A.M. Bayen, K.~Schechtner, and M.C. Gonzalez.
\newblock Understanding road usage patterns in urban areas.
\newblock {\em Nature Scientific Reports}, 2, 2012.

\bibitem{whaiduzzaman_2014}
M.~Whaiduzzaman, M.~Sookhak, A.~Gani, and R.~Buyya.
\newblock A survey on vehicular cloud computing.
\newblock {\em Journal of Network and Computer Applications}, 40:325--344,
  2014.

\bibitem{zhang_2012}
W.~Zhang, M.~Kamgarpour, D.~Sun, and C.J. Tomlin.
\newblock A hierarchical flight planning framework for air traffic management.
\newblock {\em Proceedings of the IEEE}, 100(1):179--194, 2012.

\end{thebibliography}

\end{document}